\documentclass[twocolumn,showpacs,preprintnumbers]{revtex4}
\usepackage{graphicx,epsfig,amsmath,amsfonts}

\begin{document}

\title{Nonlinear Lattice Dynamics of Bose-Einstein Condensates}
\author{Mason A.\ Porter}
\affiliation{School of Mathematics and Center for Nonlinear Science, School of 
Physics \\ Georgia Institute of Technology, Atlanta GA 30332-0160, USA}
\author{R.\ Carretero-Gonz\'alez}
\affiliation{Department of Mathematics \& Statistics and Nonlinear Dynamical 
Systems Group (http://nlds.sdsu.edu),\\
San Diego State University, San Diego, CA 92182-7720, USA}
\author{P. G.\ Kevrekidis}
\affiliation{Department of Mathematics \& Statistics, University of Massachusetts,
Amherst MA 01003-4515, USA}
\author{Boris A.\ Malomed}
\affiliation{Department of Interdisciplinary Studies, Faculty of Engineering, 
Tel Aviv University, Tel Aviv 69978, Israel}

\begin{abstract}
The Fermi-Pasta-Ulam (FPU) model, which was proposed 50 years ago
to examine thermalization in non-metallic solids and develop
``experimental'' techniques for studying nonlinear problems,
continues to yield a wealth of results in the theory and
applications of nonlinear Hamiltonian systems with many degrees of
freedom. Inspired by the studies of this seminal model,
solitary-wave dynamics in lattice dynamical systems have proven
vitally important in a diverse range of physical 
problems---including energy relaxation in solids, denaturation of 
the DNA double strand, self-trapping of light in arrays of 
optical waveguides, and Bose-Einstein condensates (BECs) in 
optical lattices. BECS, in particular, due to their widely ranging and 
easily manipulated dynamical apparatuses---with one to three 
spatial dimensions, positive-to-negative tuning of the 
nonlinearity, one to multiple components, and numerous 
experimentally accessible external trapping potentials---provide one 
of the most fertile grounds for the analysis of 
solitary waves and their interactions. In this paper, we review 
recent research on BECs in the presence of deep periodic potentials, which 
can be reduced to nonlinear chains in appropriate circumstances. These 
reductions, in turn, exhibit many of the remarkable 
nonlinear structures (including solitons, intrinsic localized 
modes, and vortices) that lie at the heart of the nonlinear science 
research seeded by the FPU paradigm.
\end{abstract}

\date{Submitted to {\em Chaos}, September 2004}
\maketitle

\vspace{2mm}

\textbf{
The Fermi-Pasta-Ulam (FPU) model was formulated in 1954 in
an attempt to explain heat conduction in non-metallic lattices and
develop ``experimental'' (computational) methods for research on
nonlinear dynamical systems \cite{fpugen}. Further studies of this
problem ten years later led to the first analytical description of
solitons (using the Korteweg-de Vries equation, which is a
continuum approximation of the discrete FPU system), which have since 
become one of the most fundamental paradigms of nonlinear
science. These nonlinear waves occur ubiquitously in rather diverse 
physical situations ranging from water waves to plasmas, optical fibers, 
superconductors (long Josephson junctions), quantum field
theories, and more. Over the past several years, the study of
solitons and coherent structures in Bose-Einstein condensates
(BECs) has come to the forefront of experimental and theoretical
efforts in soft condensed matter physics, drawing the attention
of atomic and nonlinear physicists alike. Observed experimentally 
for the first time in 1995 in vapors of sodium and rubidium
\cite{becna,becrub}, a BEC---a macroscopic cloud of
coherent quantum matter---is attained when 
($10^{3}$ to $10^{6}$) atoms, confined in
magnetic traps, are optically and evaporatively cooled to a
fraction of a microkelvin. The macroscopic behavior of BECs near
zero temperature is modeled very well by the Gross-Pitaevskii
equation (a time-dependent nonlinear Schr\"{o}dinger
equation with an external potential), which admits a wide range
of coherent structure solutions. Especially attractive is that 
experimentalists can now engineer a wide variety of external trapping
potentials (of either magnetic or optical origin) confining the 
condensate. As a key example, we focus on BECs loaded into
deep, spatially periodic optical potentials, effectively 
splitting the condensate into a chain of linearly-interacting, nonlinear
droplets, the dynamics of which is accurately characterized by
nonlinear lattice models. This paper highlights some of the
quasi-discrete nonlinear dynamical structures in BECs 
reminiscent of the discoveries that originated from the FPU model.}

\section{Introduction} \label{intro}

One of the most important nonlinear problems, whose origin dates
back to the early 20th century, concerns the conduction of heat in
dielectric crystals. As early as 1914, Peter Debye suggested that
the finite thermal conductivity of such lattices is due to the
nonlinear interactions among lattice vibrations (i.e.,
phonon-phonon scattering) \cite{norm}. To understand the process
of thermalization---which refers to how and to what extent
energy is transported from coherent modes and macroscopic scales
to internal, microscopic ones \cite{friep1}---and to develop
computational techniques for studying nonlinear dynamical systems,
Fermi, Pasta, and Ulam (FPU) posed the following question in 1954: 
How long does it take for long-wavelength oscillations to transfer 
their energy into an equilibrium distribution in a one-dimensional
string of nonlinearly interacting particles?  This question has since 
spawned a diverse array of activities attempting to answer it and 
fomented a strong impetus to research in topics such as soliton 
theory, discrete lattice dynamics, and KAM theory. Furthermore, these fronts 
remain active research topics \cite{friep1,friep2,friep3,friep4,Lepri}.

Before the FPU work, it was commonly assumed that high-dimensional
Hamiltonian systems behave ergodically in the sense that a smooth 
initial energy distribution should quickly relax until it is ultimately 
distributed evenly among all of the system's modes (that is, thermalization 
should occur).  To explicitly verify
this fundamental hypothesis of statistical physics, FPU constructed a
one-dimensional dynamical lattice, with $N$ identical particles which 
interact according to an anharmonic repulsive force.



Running numerical simulations on the computers available in the
early fifties, FPU observed that the lattice
did not relax to thermal equilibrium, contrary to everybody's 
expectations \cite{fpu55,norm,lich,ablo,fpugen}.  An especially
striking observation was a beating effect, in the form of a 
near-recurrence of the initial long-scale configuration, which 
reappeared after a large number of oscillations involving 
short-scale modes.  In this manner, more than $97\%$ of the 
energy returned to the initial mode. Moreover, this finding was 
robust with respect to variations in the total number of particles 
and particular choice of the (power-law) anharmonicity in the 
interaction between them.

Motivated by this study, Zabusky and Kruskal considered a
continuum version of the model, showing that the dynamics of
small-amplitude, long-wavelength perturbations obeys [on the
timescale $\sim $ $($wavelength$)^{3}$] the Korteweg-de Vries
(KdV) equation. They subsequently introduced the concept of 
solitons (solitary waves) in terms of the KdV equation
\cite{ablo,zab,norm}. The explanation for the lack of 
thermalization is that the energy gets concentrated in robust
coherent structures (the solitons), which interact elastically 
and thus do not transfer their energy into linear lattice modes 
(phonon waves, also referred to as ``radiation") \cite{friep1}. 
The KdV equation thereby became the first example in the celebrated 
class of nonlinear partial differential equations (PDEs) that are integrable 
by means of the inverse scattering transform.  It was later followed by 
numerous other important PDEs, such as the nonlinear Schr{\"o}dinger 
equation, the sine-Gordon equation, higher dimensional 
examples (including the Kadomtsev-Petviashvili and the Davey-Stewartson 
equations) and multi-component examples (including the Manakov 
equation) \cite{JCE,Drazin}.

Since then, the study of solitons, and more general coherent structures, has
become one of the paradigms of nonlinear science \cite{whitham,ablo,ist}.
Such dynamical behavior occurs in a wide variety of physical 
systems---including (to name just a few examples) nonlinear optics, 
fluid mechanics, plasma physics, and quantum field theory. Over the 
past several years, the impact of solitary-wave dynamics has been 
especially significant in the study of Bose-Einstein condensates 
(BECs) \cite{pethick,stringari}. In this short review,
we focus on this application and, in particular, its description in
physically appropriate cases in terms of dynamical lattices.

The rest of this paper is organized as follows. In Section
\ref{fpuintro}, we define the FPU problem and briefly survey its
mathematical properties. In Section \ref{becintro}, we provide an
introduction to Bose-Einstein condensates and their
solitary wave solutions. We consider, in particular, BECs loaded
into optical lattices (OLs), and use a Wannier-function expansion to 
derive a dynamical lattice model describing this system.  In
appropriate limits, this leads to a (generally,
multiple-component) discrete nonlinear Schr\"{o}dinger (DNLS)
equation \cite{alf}. In Section \ref{tail}, we examine the regime
in which a deep, spatially periodic OL potential effectively fragments
the BEC into a chain of weakly interacting droplets. The resulting
model, which consists of a Toda lattice with on-site potentials,
produces self-localized modes that may be construed as solitons
of the underlying BEC. Section \ref{conc} concludes the paper,
discussing a number of future directions.

\section{The FPU Problem} \label{fpuintro}

The FPU model consists of a chain of particles connected by nonlinear
springs. The constitutive law of the model, i.e., the relation
between the interaction force $F$ and the distance $y$ between
adjacent atoms in the chain (with only nearest-neighbor interactions 
postulated), was taken to be $F(y)=-[y+G(y)]$. FPU considered 
three different functional forms of $G(y)$: quadratic, cubic, and 
piecewise linear \cite{fpu55}.



In the case of a cubic force law \cite{lich}, $F(y)=-(y+3\beta y^{3})$, where 
$\beta $ is an effective anharmonicity coefficient, one may write
\begin{equation}
  \begin{array}{rcl}
\displaystyle\ddot{y}_{j} & = &
\displaystyle\frac{y_{j+1}-2y_{j}+y_{j-1}}{h^{2}}\left\{
1+\frac{\beta }{h^{2}}\left[ (y_{j+1}-y_{j})^{2}\right. \right.
\\[2ex]
& + & \displaystyle\left. \left.
(y_{j}-y_{j-1})^{2}+(y_{j+1}-y_{j})(y_{j}-y_{j-1})\right]
\vphantom{\frac{\beta}{h^2}}\right\} \,,\end{array}~ \label{dstring}
\end{equation}
which is supplemented with fixed boundary conditions, $y_{0}=y_{N}=0$.  
FPU \cite{fpu55,ablo} used the initial condition 
$y_{j}(0)=\sin (j\pi /N)$ and $\dot{y}_{j}(0)=0$.  This form (\ref{dstring}) 
of the FPU chain is obtained by discretizing the continuous nonlinear string,
\begin{equation}
  \frac{\partial ^{2}y}{\partial t^{2}}-\frac{\partial
^{2}y}{\partial x^{2}}\left[ 1+3\beta \left( \frac{\partial
y}{\partial x}\right) ^{\!2}\right] =0  \label{string}
\end{equation}
with the following approximations to the continuum derivatives:
\begin{equation}
  \begin{array}{rcl}
\displaystyle\frac{\partial y}{\partial x} & \equiv  & \displaystyle\left\{
\frac{y_{j+1}-y_{j}}{h}\,,\,\frac{y_{j}-y_{j-1}}{h}\right\},  \\[2.5ex]
\displaystyle\frac{\partial ^{2}y}{\partial x^{2}} & \equiv  &
\displaystyle\frac{y_{j+1}-2y_{j}+y_{j-1}}{h^{2}}.\end{array} \label{disc}
\end{equation}
Here, $y_{j}$ is the displacement of the $j$-th particle from its
equilibrium position, $h\equiv L/N$ is the normalized spacing (the
distance between the particles), $L$ is the string's length, and $N$
is the number of particles, which FPU took to be $16$, $32$, or $64$
in their calculations.  (Using equation (\ref{dstring}), the onset of
resonance overlaps was studied in the FPU problem in the first ever
application of Chirikov's overlap criterion \cite{lich,iz}.)

With an appropriate scaling 
($t\rightarrow ht$, $y\rightarrow hy\sqrt{3/\beta }$), the FPU chain can also 
be written
\begin{align}
 \ddot{y}_j &= (y_{j+1} - y_j) - (y_j - y_{j-1}) \notag \\
&\quad + \frac{1}{3}\left[(y_{j+1} - y_j)^3 - (y_j - y_{j-1})^3\right]\,. 
\label{resc}
\end{align}
Using the continuum field variable $u$ \cite{Driscoll:76},
\begin{equation}
  u\equiv -\frac{y_{t}}{2h}+\frac{1}{2}\int_{0}^{y_{x}}(1+h^{2}\eta
^{2})^{1/2}d\eta,
\end{equation}
Eq.\ (\ref{string}) yields, to lowest order in $h$, the modified Korteweg-de
Vries (mKdV) equation (with $\tau \equiv h^{3}t/24$, $\xi \equiv x-ht$),
\begin{equation}
u_{\tau }+12u^{2}u_{\xi }+u_{\xi \xi \xi }=0\,,
\end{equation}
which is further reduced to the KdV equation proper via the
Miura transformation \cite{norm,ablo,whitham,ist}.

One can also derive the KdV equation directly from an FPU chain
with a quadratic anharmonicity in the inter-particle interaction
\cite{ablo,fpu55,zab}, $F(y)=-(y+\alpha y^{2})$. In the latter
case, the discrete FPU model takes the form
\begin{equation}
\ddot{y}_{j}=\frac{y_{j+1}-2y_{j}+y_{j-1}}{h^{2}}\left( 1+\alpha
\frac{y_{j+1}-y_{j-1}}{h}\right) \,.  \label{disc2}
\end{equation}To study the near-recurrence phenomenon in 
Eq.\ (\ref{disc2}), Zabusky and Kruskal \cite{zab,norm} derived 
its continuum limit ($h \longrightarrow 0$, $Nh \longrightarrow 1$),
\begin{equation}
  y_{tt}=y_{xx}+\varepsilon
y_{x}y_{xx}+\frac{1}{12}h^{2}y_{xxxx}+O(\varepsilon h^2,h^{4})\,,
\label{PDE}
\end{equation}
where $\varepsilon \equiv 2\alpha h$.

Unidirectional asymptotic solutions to Eq.\ (\ref{PDE}) are constructed
with \cite{ablo}
\begin{equation}
  y \sim \phi (\xi ,\tau )\,,\quad \xi \equiv x-t\,,\quad \tau \equiv
\frac{1}{2}\varepsilon t,
\end{equation}
where the function $\phi $ obeys the equation
\begin{equation}
\phi _{\xi \tau }+\phi _{\xi }\phi _{\xi \xi }+\delta ^{2}\phi _{\xi \xi \xi
\xi }+O\left( h^{2},h^{4}\varepsilon ^{-1}\right) \,=0\,,  \label{xi-tau}
\end{equation}
for small $\delta ^{2}\equiv h^{2}\varepsilon ^{-1}/12$. 
Finally, with $u\equiv \phi _{\xi }$, one obtains 
the KdV equation from  Eq.\ (\ref{xi-tau}) \cite{krusk}:
\begin{equation}
  u_{\tau }+uu_{\xi }+\delta ^{2}u_{\xi \xi \xi }\,=0\,. \label{kdv}
\end{equation}
Eq.\ (\ref{kdv}) has solitary-wave solutions of the form
\begin{equation}
u(x,t)=2\kappa ^{2}\mbox{sech}^{2}\left[ \kappa (x-4\kappa
^{2}t-x_{0})\right] \,,  \label{solit}
\end{equation}
with constants $\kappa $ and $x_{0}$. [Note that a more 
rigorous derivation of the KdV equation from the FPU chain 
(as a fixed point of a renormalization process)
has recently been developed \cite{friep1}.]

Although the solitary-pulse solution (\ref{solit}) has been well-known
since the original paper by Korteweg and de Vries, it was the
paper by Zabusky and Kruskal \cite{zab} that revealed the
particle-like behavior of the pulses in numerical simulations.  (The term 
``soliton" was coined in that paper to describe them.)  Since then, solitons 
have become ubiquitous, as their study has yielded vital insights into
numerous physical problems \cite{whitham,ablo,ist,nwm,Drazin,JCE}. In the
next section, we discuss their importance to Bose-Einstein
condensation \cite{stringari,pethick}.

It is remarkable that even today, 50 years from the original
derivation \cite{fpu55}, FPU chains are themselves still studied
as a means of understanding a variety of nonlinear phenomena.
Recent studies focus not only on the model's solitary-wave 
solutions and their stability \cite{friep1,friep2,friep3,friep4,rink1},
but also on its thermodynamic properties and connections with the 
Fourier law of heat conductivity \cite{Lepri},
its dynamical systems/invariant manifold aspects \cite{rink2,rink3}, 
and its connections with weak turbulence theory \cite{Biello}.

To conclude this section, we remark that the most natural discrete
model that has been derived from the NLS (or GP) equation for a
soliton train is the Toda lattice \cite{kaup} (see also the details
discussed below).  However, the leading-order nonlinear truncation of
the latter lattice equation once again yields the FPU model.
Conversely, one can approximate the FPU chain by the NLS equation in
the high-frequency limit \cite{berkol,lich}.  The validity of this
approximation varies with time due to energy exchange between modes.

\section{Bose-Einstein Condensation} \label{becintro}

At low temperatures, bosonic particles in a dilute gas can reside 
in the same quantum (ground) state, forming a Bose-Einstein 
condensate (BEC) \cite{pethick,stringari,ketter,edwards}. Seventy years 
after they were first predicted theoretically, BECs were finally 
observed experimentally in 1995 in vapors of rubidium and 
sodium \cite{becrub,becna}. In these experiments, atoms were loaded 
into magnetic traps and evaporatively cooled to temperatures on the 
order of a fraction of a microkelvin. To record the properties of 
the BEC, the confining trap was then switched off, and the expanding 
gas was optically imaged \cite{stringari}. A sharp peak in the velocity 
distribution was observed below a critical temperature, indicating that 
condensation had occurred.

Under experimental conditions, BECs are inhomogeneous, so 
condensation can be observed in both momentum and coordinate
space. The number of condensed atoms $\mathcal{N}$ ranges from
several thousand (or less) to several million (or more). The
magnetic traps confining BECs are usually well-approximated by
harmonic potentials. There are two characteristic length scales.
One is the harmonic oscillator length,
$a_{\mathrm{ho}}=\sqrt{\hbar /(m\omega _{\mathrm{ho}})}$
(which is, typically, on the order of a few microns), where $m$ is the
atomic mass and $\omega _{\mathrm{ho}}=(\omega _{x}\omega
_{y}\omega _{z})^{1/3}$ is the geometric mean of the trapping
frequencies. The second scale is the mean healing length, $\chi
=1/\sqrt{8\pi |a|\bar{n}}$, where $\bar{n}$ is the mean density of
the atoms, and $a$, the (two-body) $s$-wave scattering length, is
determined by collisions between atoms
\cite{pethick,stringari,kohler,baiz}. Interactions between atoms
are repulsive when $a>0$ and attractive when $a<0$. The length
scales in BECs should be compared to those in condensed media like
superfluid helium, in which the effects of inhomogeneity occur on
a microscopic scale fixed by the interatomic distance \cite{stringari}.

With two-body collisions described in the mean-field
approximation, a dilute Bose-Einstein gas is very accurately
modeled by the cubic nonlinear Schr\"{o}dinger equation (NLS)
with an external potential [i.e., by the so-called
Gross-Pitaevskii (GP) equation]. An important case is that of 
cigar-shaped BECs, which are tightly confined in two transverse 
directions (with the radius on the order of the healing length) 
and quasi-free in the longitudinal 
dimension \cite{salasnich,Perez-Garcia:98,Salasnich:02a,Boris:03,Lieb:03}.
In this regime, one employs the 1D limit of the 3D mean-field
theory, generated by averaging in the transverse plane (rather
than the 1D mean-field theory per se, which would be appropriate
were the transverse dimension on the order of the atomic size
\cite{salasnich,Perez-Garcia:98,Salasnich:02a,Boris:03,Lieb:03}).

The original GP equation, describing the BEC near zero temperature, is
\cite{Gross:61,Pitaevskii:61,stringari,kohler,baiz}
\begin{equation}
  i\hbar {\Psi }_{t}=\left( -\frac{\hbar ^{2}\nabla ^{2}}{2m}+g_{0}|\Psi
|^{2}+{\mathcal{V}}(\vec{r})\right) \Psi ,  \label{GPE}
\end{equation}
where $\Psi =\Psi (\vec{r},t)$ is the condensate wave function
normalized to the number of atoms $\mathcal{N}$, 
${\mathcal{V}}(\vec{r})$ is the external potential, and the
effective interaction constant is \cite{stringari} $g_{0}=[4\pi
\hbar ^{2}a/m][1+O(\zeta ^{2})]$, where $\zeta \equiv
\sqrt{|\Psi |^{2}|a|^{3}}$ is the dilute-gas parameter. The
resulting normalized form of the 1D equation is
\cite{salasnich,Perez-Garcia:98,Salasnich:02a,Boris:03,Lieb:03}
\begin{equation}
  i\psi _{t}=-\frac{1}{2}\psi _{xx}+g|\psi |^{2}\psi +V(x)\psi ,  \label{nls3}
\end{equation}
where $\psi $ and $V$ are, respectively, the rescaled 1D wave 
function (a result of averaging in the transverse directions) and 
external potential.  The rescaled self-interaction parameter $g$ is tunable 
(even its sign), because the scattering length $a$ can be adjusted using 
magnetic fields in the vicinity of a Feshbach resonance \cite{fesh,FRM}.

\subsection{BECs in Optical Lattices and Superlattices}

BECs can be loaded into optical lattices (or superlattices, which
are small-scale lattices subjected to a long-scale periodic
modulation), which are created experimentally as interference
patterns of counter-propagating laser
beams \cite{Anderson:98,g2b,g2c,g3,g4,Smerzi:01,g6,g7}. Over the
past several years, a vast research literature has developed
concerning BECs in such potentials
\cite{anderson,mott,lattice,morsch,space2,tight,quasibec,anamaria,mapbecprl,super,bronski,pk1,pk2,pk3},
as they are of considerable interest both experimentally and
theoretically. Among other phenomena, they have been used to study
Josephson effects \cite{anderson}, squeezed states \cite{squeeze},
Landau-Zener tunneling and Bloch oscillations \cite{morsch},
controllable condensate splitting \cite{super}, and the transition
between superfluidity and Mott insulation at both the classical
\cite{smerzi,cata} and quantum \cite{mott} levels. Moreover, with
each lattice site occupied by one alkali atom in its ground state,
BECs in optical lattices show promise as a register in a quantum
computer \cite{porto,voll}.

With the periodic potential $V(x)=V(x+L)$, one may examine stationary
solutions to (\ref{nls3}) in the form
\begin{equation}
  \psi (x,t)=R(x)\exp \left( i\left[ \theta (x)-\mu t\right] \right) \,,
\label{coher}
\end{equation}
where $\mu$, the BEC's chemical potential, is determined by the
number of atoms in the BEC; it is positive for repulsive BECs and 
can assume either sign for attractive BECs. Using the relation
$d\theta /dx=c/R^{2}$ (``angular momentum'' conservation), one derives a
parametrically forced Duffing-oscillator equation for the
amplitude function \cite{mapbecprl,mapbec,bronski,bronskiatt,bronskirep,alf2},
\begin{equation}
  R''-\frac{c^{2}}{R^{3}}+\mu R-gR^{3}-V(x)R=0\,,  \label{duff}
\end{equation}
where $R'' \equiv d^2R/dx^2$.

Equation (\ref{duff}) admits both localized and spatially extended 
solutions.  Supplemented with appropriate boundary conditions, it yields 
both bright and dark solitons, which correspond, respectively, to 
localized humps on the zero background, and localized dips in a 
finite-density background.  These states are similar to the bright 
and dark solitons in nonlinear optics; they are stable,
respectively, in attractive and repulsive 1D BECs \cite{alf2,band}.

When $V(x)$ is spatially periodic, the bright solitons resemble gap solitons, 
which are supported by Bragg gratings in nonlinear optical systems. In BECs, 
they have been observed in two situations: (1) the small-amplitude limit, with
the value of $\mu $ close to forbidden zones (``gaps'') of
the underlying linear Schr\"{o}dinger equation with a periodic
potential \cite{kon,sakaguchi}; and (2) in the
\textit{tight-binding approximation} (discussed below), for which
the continuous NLS equation can be replaced by its discrete counterpart---the 
so-called discrete nonlinear Schr\"{o}dinger (DNLS) equation
\cite{abd}. In the latter context, the strongly localized
solutions are known as intrinsic localized modes (ILMs) or
discrete breathers. Spatially extended wave functions with
periodic or quasi-periodic $R(x)$, which may be either resonant or
non-resonant with respect to the periodic potential $V(x)$, are
known as \textit{modulated amplitude waves} and have been shown
to be stable (against arbitrary small perturbations) in some cases
\cite{bronski,bronskiatt,bronskirep,mapbecprl,mapbec,super}.

\subsection{Lattice Dynamics}

In the presence of a strong optical lattice, the GP equation (\ref{nls3})
can be reduced to the DNLS equation \cite{alf,alf2,kef}. To 
justify this approximation, the wave function is expanded in 
terms of a set of Wannier functions localized near the minima of
the potential wells. 

The eigenvalue problem associated with the linear part of 
Eq.\ (\ref{nls3}) is
\begin{equation}
  -\varphi _{k,\alpha }^{\prime \prime }+V(x)\varphi _{k,\alpha }=E_{\alpha
}(k)\varphi _{k,\alpha }\,,
\end{equation}
where $\varphi _{k,\alpha }$ can be expressed in terms of 
Floquet-Bloch functions, $\varphi _{k,\alpha }=e^{ikx}u_{k,\alpha
}(x)$, with $u_{k,\alpha }(x)=u_{k,\alpha }(x+L)$ and $\alpha $
indexing the energy bands, so that $E_{\alpha }(k)=E_{\alpha
}(k+[2\pi /L])$ \cite{ashcroft}. The energy is represented using
Fourier series,
\begin{equation}
  E_{\alpha }(k)=\sum_{n=-\infty }^{\infty }\hat{\omega}_{n,\alpha
}e^{iknL}\,,\quad \hat{\omega}_{n,\alpha }=\hat{\omega}_{-n,\alpha
}=\hat{\omega}_{n,\alpha }^{\ast }\,,  \label{four}
\end{equation}
where the asterisk denotes complex conjugation and
\begin{equation}
\hat{\omega}_{n,\alpha }=\frac{L}{2\pi }\int_{-\pi /L}^{\pi /L}E_{\alpha
}(k)e^{-iknL}dk\,.
\end{equation}

Although the Floquet-Bloch functions provide a complete
orthonormal basis, it is more convenient to utilize Wannier
functions (also indexed by $\alpha$),
\begin{equation}
  w_{\alpha }(x-nL)=\sqrt{\frac{L}{2\pi }}\int_{-\pi /L}^{\pi /L}\varphi
_{k,\alpha }(x)e^{-inkL}dk\,,
\end{equation}
which are centered about $x=nL$ ($n\in \mathbb{Z}$).  The Wannier functions 
constitute a complete orthonormal basis with respect to both $n$ and 
$\alpha$. One can also guarantee that the Wannier functions are real 
by conveniently choosing phases of the Floquet-Bloch functions.

Given the orthonormality of the Wannier function basis, any
solution of (\ref{nls3}) can be expanded in the form
\begin{equation}
  \psi (x,t)=\sum_{n,\alpha }c_{n,\alpha }(t)w_{n,\alpha }(x)\,,
\end{equation}
with coefficients satisfying a DNLS equation with long-range interactions,
\begin{equation}
\begin{array}{rcl}
\displaystyle i\frac{d c_{n,\alpha }}{d t} & = & \displaystyle\sum_{n_{1}}c_{n_{1},\alpha }\hat{\omega}_{n-n_{1},\alpha } \\[3ex]
& + & \displaystyle g\sum_{\alpha _{1},\alpha _{2},\alpha
_{3}}\sum_{n_{1},n_{2},n_{3}}c_{n_{1},\alpha _{1}}^{\ast
}c_{n_{2},\alpha _{2}}c_{n_{3},\alpha _{3}}W_{\alpha ,\alpha
_{1},\alpha _{2},\alpha _{3}}^{n,n_{1},n_{2},n_{3}}\,,\end{array}
\label{over}
\end{equation}
as was illustrated in \cite{alf}. In (\ref{over}),
\begin{equation}
  W_{\alpha ,\alpha _{1},\alpha _{2},\alpha
_{3}}^{n,n_{1},n_{2},n_{3}}=\int_{-\infty }^{\infty }w_{n,\alpha
}w_{n_{1},\alpha _{1}}w_{n_{2},\alpha _{2}}w_{n_{3},\alpha _{3}}dx\,.
\label{over_int}
\end{equation}
Because the Wannier functions are real, the integral in (\ref{over_int}) 
is symmetric with respect to all permutations of both 
$(\alpha ,\alpha _{1},\alpha _{2},\alpha_{3})$ and $(n,n_{1},n_{2},n_{3})$.

Although Eqs.\ (\ref{over}) are intractable as written, several
important special cases can be studied \cite{alf}. The
nearest-neighbor coupling approximation is valid when
$|\hat{\omega}_{1,\alpha }|\gg |\hat{\omega}_{n,\alpha }|$ for
$n>1$. More generally, one can assume that the Fourier
coefficients in (\ref{four}) decay rapidly beyond a finite number
of harmonics. This simplifies the linear term in (\ref{over}). 
Additionally, because $w_{n,\alpha }$ is localized
about its center at $x=nL$, it is sometimes reasonable to assume
that the coefficients satisfying $n=n_{1}=n_{2}=n_{3}$ dominate
$W_{\alpha ,\alpha _{1},\alpha _{2},\alpha
_{3}}^{n,n_{1},n_{2},n_{3}}$, so that the others may be neglected.
In the nearest-neighbor regime, this implies that
\begin{align}
\begin{array}{rcl}
\displaystyle i\frac{d{c}_{n,\alpha}}{dt} & = & \hat{\omega}_{0,\alpha }c_{n,\alpha
}+\hat{\omega}_{1,\alpha }(c_{n-1,\alpha }+c_{n+1,\alpha }) \\[2ex]
& + & \displaystyle g\sum_{\alpha _{1},\alpha _{2},\alpha
_{3}}W_{\alpha ,\alpha _{1},\alpha _{2},\alpha _{3}}c_{n,\alpha
_{1}}^{\ast }c_{n,\alpha _{2}}c_{n,\alpha
_{3}}\,,\label{nn}\end{array}
\end{align}
where $W_{\alpha,\alpha _{1},\alpha _{2},\alpha _{3}}\equiv W_{\alpha ,\alpha
_{1},\alpha _{2},\alpha _{3}}^{n,n,n,n}$ is independent of $n$.
This leads to the tight-binding model,
\begin{equation}
  \begin{array}{rcl}
\displaystyle i\frac{d{c}_{n,\alpha}}{dt} & = & \hat{\omega}_{0,\alpha }c_{n,\alpha
}+\hat{\omega}_{1,\alpha }(c_{n-1,\alpha }+c_{n+1,\alpha }) \\[2ex]
& + & \displaystyle gW_{\alpha ,\alpha,\alpha,\alpha}
|c_{n,\alpha }|^{2}c_{n,\alpha
}\,,\label{tight}\end{array}\end{equation}
in the single-band approximation. Typically, this approximation
is valid if the height of the barrier between potential wells
is large and if the wells are well-separated. While this intuition may
be generally true, the Wannier function reduction 
provides a systematic tool that
can establish the validity of the approximation on a case by case basis
(by determining the overlap coefficients); see, e.g., \cite{alf}
for specific examples.
Including next-nearest-neighbor 
coupling in this regime allows one to study interactions between 
intrasite and intersite nonlinearities. 

\begin{figure}[th]
\centerline{
\includegraphics[width=8.25cm,angle=0]{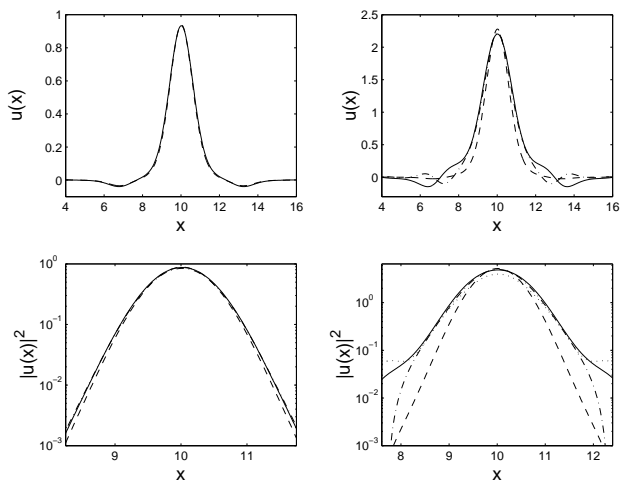}
}
\caption{ Comparison of the
lattice reconstructed solution in the tight-binding (dashed line)
and the 3-band (dash-dotted line) approximation with the exact
solution (solid line). The comparison is performed for
$V(x)=-5 \cos(2 x)$ and different chemical potentials: 
$\mu=1.5$ (left panels) and $\mu=-1.5$ (right panels).  
The bottom (semi-log) panels show the result 
of dynamical time-evolution of the tight-binding 
approximation with the dotted line.
As time evolves (we depict snapshots at $t \approx 50$), 
the latter can be seen to approach the shape of the exact 
solution (in the left panel, these plots cannot be distinguished)
and to match its asymptotic form, possibly shedding small wakes 
of low-amplitude wave radiation in the process (see, for example, 
the bottom right panel).} \label{wff}
\end{figure}

In more general situations in which single-band descriptions are
inadequate because of resonant, nonlinearity-induced interband
interactions, one can simplify (\ref{nn}) by assuming that
$\hat{\omega}_{1,\alpha }\equiv
\hat{\omega}_{1,\alpha }(\epsilon )=O(\epsilon )$ as
$\epsilon \rightarrow 0$. Applying the phase shift 
$c_{n,\alpha }(t)=\exp [i\hat{\omega}_{0,\alpha
}t]\tilde{c}_{n\alpha }(t)$, so that the nonlinear terms consist
of oscillatory exponents, and supposing that $\tilde{c}_{n,\alpha
}(0)$ is sufficiently small and may be ignored, we conclude that,
on the time scale $\sim 1/\epsilon$, the exponents oscillate
rapidly except when $\alpha =\alpha _{2}$, $\alpha _{1}=\alpha
_{3} $ or $\alpha =\alpha _{3}$, $\alpha _{1}=\alpha _{2}$. Through
time-averaging, one obtains
\begin{equation}
\begin{array}{rcl}
\displaystyle i\frac{d\tilde{c}_{n,\alpha }}{dt} & = & \hat{\omega}_{1,\alpha
}(\tilde{c}_{n-1,\alpha }+\tilde{c}_{n+1,\alpha }) \\[2ex]
\displaystyle & + & \displaystyle g\sum_{\alpha _{1}}W_{\alpha
,\alpha _{1}}|\tilde{c}_{n,\alpha _{1}}|^{2}\tilde{c}_{n,\alpha
_{1}}\,,\label{dn}\end{array}\end{equation}
where 
$W_{\alpha,\alpha _{1}}\equiv W_{\alpha ,\alpha _{1},\alpha ,\alpha _{1}}$
describes the interband interactions. Equation (\ref{dn}) is a
vector DNLS with cross-phase-modulation nonlinear coupling. 
An example of the implementation of this method is illustrated in 
Fig.\ \ref{wff}.  More generally, the advantage of the approach of 
Ref. \cite{alf} is that, given the explicit form of the potential, 
the relevant coefficients can be computed and the appropriate reduced 
(single-band or multiple-band) model can be derived to the desired 
level of approximation.

At this stage, one can study the ILMs of Eqs.\ (\ref{tight}) or
(\ref{dn}) and use the Wannier function expansion to reconstruct
the solution of the original GP equation \cite{alf}. 
This approach has been successfully used in a variety of 
applications and for different localized states present in
the lattice---including bright, dark, and discrete-gap solitons, as well as
breathers \cite{abd,smer,pk3,alf2}.  Other phenomena, such as discrete
modulational instabilities, have also been studied \cite{kef}.
More specifically, one of the most successful implementations
of discrete NLS equations (and variants thereof) in this context included
the quantitative prediction that its modulational stability
analysis determined the threshold of a dynamical instability
of the condensate (the so-called classical superfluid-insulator
transition of \cite{smerzi}). These predictions were subsequently 
verified quantitatively by the experimental measurements of Ref. \cite{cata}.

\section{Soliton-soliton tail-mediated interactions and the Toda lattice} 
\label{tail}

\begin{figure}[th]
\centerline{
\includegraphics[width=8.25cm,angle=0]{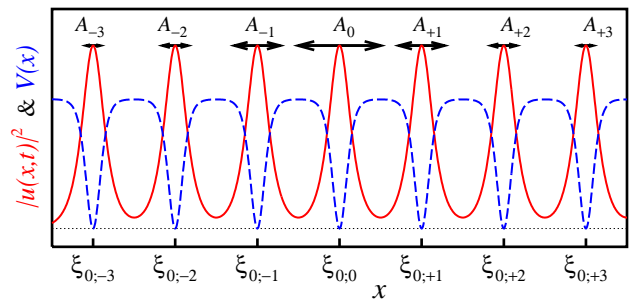}
} 
\caption{A quasi-1D condensate (solid line) in a deep
periodic optical-lattice potential (dashed line). The condensate
is effectively described as a chain of coupled solitons whose
positions follow a Toda lattice with on-site potentials
[Eq.\ (\ref{LDE})]. Using the oscillating ansatz 
(\protect\ref{osc-ansatz1}), where the $j$-th soliton is forced 
to oscillate with amplitude $A_{j}$, one can further reduce the 
dynamics to a second-order recurrence relationship between neighboring 
amplitudes [Eq.\ (\ref{recurrence})].} \label{train}
\end{figure}

Recent advances in trapping techniques allow the
generation of bright solitons and chains of bright solitons in
effectively 1D attractive condensates
\cite{Strecker:02,Khawaja:02,Strecker:03,Carr:02}. In this
section, we consider the collective motion of a chain of bright
solitons. We focus, in particular, on the dynamics of attractive
BECs trapped in a deep optical lattice (OL) that renders a 1D
attractive condensate into a chain of interacting solitons (see
Fig.\ \ref{train}). 

Consider a BEC loaded into an OL
potential produced by the interference pattern of multiple
counter-propagating laser beams
\cite{Anderson:98,g2b,g2c,g3,g4,Smerzi:01,g6,g7}. In
principle, it is possible to design various optical trap profiles 
(essentially at will) by appropriately superimposing interference patterns.
For the purposes of this exposition, we adopt an OL profile, with a tunable 
inter-well separation, given by the Jacobi elliptic-sine function,
\begin{equation}
  V(x)=V_{0}\,\mathrm{sn}^{2}(x;k)\,,  \label{Vsn}
\end{equation}
where $V_{0}$ is the strength of the OL. The elliptic modulus $k$
allows one to tune the separation between consecutive wells, $r \equiv \xi
_{0;j+1}-\xi _{0;j}=2K(k)$, where $\xi _{0;j}=2j\,K(k)$ is the position of
the $j$-th well and $K(k)$ is the complete elliptic integral of the first
kind.

The stability properties of BECs in the optical lattice potential
(\ref{Vsn}) have been recently studied \cite{bronski, bronskirep,
bronskiatt}. Here, we are interested in the case of large
separation between the wells ($k\simeq 1$), when the BEC is
effectively fragmented into a chain of nearly identical solitons
with tail-mediated interactions, subject to the action of an
effective on-site potential due to the OL. For a large set of
parameter values, the system can be reduced to a Toda lattice
\cite{Toda:book}, as was illustrated in Refs. \cite{promislow,ric2}.
The reduction involves two steps.  

First, the interaction between the $j$-th soliton [at position 
$\xi_{j}(t)$] and the $j$-th well is approximated, using variational 
techniques \cite{Progress} or methods based on conserved quantities
\cite{ric2}, by
\begin{equation}
\ddot{\xi}_{j}=-V_{\mathrm{eff}}^{\prime }(\xi _{j}-\xi _{0;j})\,,  \label{xi}
\end{equation}
which describes a particle in the effective potential $V_{\mathrm{eff}}$ 
felt by the soliton. For well-separated troughs ($k\simeq 1$), the 
effective potential may be approximated by \cite{ric2}
\begin{equation}
  V_{\mathrm{eff}}^{\prime }(\xi )\approx \nu V_{0}\left(a_1\xi 
-\frac{224}{1125}\xi V_{0} + a_3\xi^{3}\right) \,,  \label{Veff_pert}
\end{equation}
where $\nu$ is the average amplitude (height) of the soliton, $a_1 = 8/15$ and 
$a_3 = -16/63$. 

Second, one treats the interaction of consecutive solitons in the 
absence of the OL. This interaction is well-studied in the context 
of optical solitons 
\cite{Karpman:81a,Karpman:81b,kaup,Gerdjikov:97,Arnold:98,Arnold:99,Progress}. 
 For identical, well-separated solitons (with the
phase difference $\pi$ between adjacent solitons), it is 
approximated by the Toda-lattice equation for the soliton positions,
\begin{equation}
\begin{array}{rcl}
\ddot{\xi}_{j} & = & T_{L}(\xi _{j-1},\xi _{j},\xi _{j+1}) \\[2ex]
& = & 8\nu^{3}\,\left( e^{-\nu(\xi _{j}-\xi _{j-1})}-e^{-\nu(\xi
_{j+1}-\xi _{j})}\right) \,.\end{array} \label{TL}
\end{equation}
Finally, after combining (\ref{TL}) with the on-site potential 
dynamics (\ref{xi}), we reduce the dynamics of a weakly coupled 
BEC in the deep OL to the Toda lattice with on-site potentials,
\begin{equation}
  \ddot{\xi}_{j}=T_{L}(\xi _{j-1},\xi _{j}\,, 
\xi_{j+1})-V_{\mathrm{eff}}^{\prime }(\xi _{j}-\xi _{0;j})\,.
\label{LDE}
\end{equation}

The Toda lattice (\ref{TL}) admits exact traveling-soliton
solutions \cite{Toda:book}. However, the effective potential in
(\ref{LDE}) breaks the lattice's translational invariance and
accommodates the existence of ILMs (breathers). To describe such
localized oscillations, we consider small vibrations about the
equilibrium state ($\xi _{j}=\xi _{0;j}$):
\begin{equation}
  \xi _{j}(t)=\xi _{0;j}+A_{j}\,\cos (\omega t)\,,  \label{osc-ansatz1}
\end{equation}
where $\omega$ is the common oscillation frequency for all solitons, and
the $j$-th soliton vibrates with an amplitude $A_{j}$ about
its equilibrium position $\xi _{0;j}$ (see Fig.\ \ref{train}). Substituting
the ansatz (\ref{osc-ansatz1}) into Eq.\ (\ref{LDE}) and discarding
higher-order modes yields a recurrence relationship between consecutive
amplitudes,
\begin{equation}
  A_{n+1}=(a+b\,A_{n}^{2})\,A_{n}-A_{n-1}\,,  \label{recurrence}
\end{equation}
where $a=2-\omega ^{2}+4 a_1 d$, $b=3 a_3 d$, $d=e^{\nu r}/(16\nu^{4})$, 
$r$ is the separation between adjacent troughs,
and $a_1$ and $a_3$ are the coefficients from the expansion
of the effective potential (\ref{Veff_pert}). Note that the
method just described is applicable to any OL profile that
reduces the dynamics of a single soliton to that of a particle
inside an effective potential given by a cubic polynomial in $\xi$. 

By defining consecutive oscillation amplitudes as $y_{n}\equiv
A_{n}$ and $x_{n}\equiv A_{n-1}$, the recurrence relationship
(\ref{recurrence}) can be cast as a 2D map,
\begin{equation}
\left\{
\begin{array}{rcl}
x_{n+1} & = & y_{n} \\[2ex]
y_{n+1} & = & (a+b\,y_{n}^{2})\,y_{n}-x_{n}\end{array}\right. \,.
\label{2DM}
\end{equation}
It is crucial to note here that the iteration index $n$ in (\ref{2DM})
corresponds to the \emph{spatial} lattice-site index for the
soliton chain. Thus, forward (backward) iteration of the 2D map
(\ref{2DM}) amounts to a right (left) shift of the spatial
position in the soliton chain. As our goal is to find spatially
localized states of the BEC chain (i.e., ILMs), we are interested
in finding solutions of Eq.\ (\ref{recurrence}) for which
$A_{n}\rightarrow 0$ as $n\rightarrow \pm \infty $ and $A_{0}>0$
(see Fig.\ \ref{train}).  That is, $(x_{n},y_{n})\rightarrow (0,0)$ as
$n\rightarrow \pm \infty $. These solutions correspond to
homoclinic connections of the origin.

\begin{figure}[th]
\centerline{
\includegraphics[width=8.25cm,angle=0]{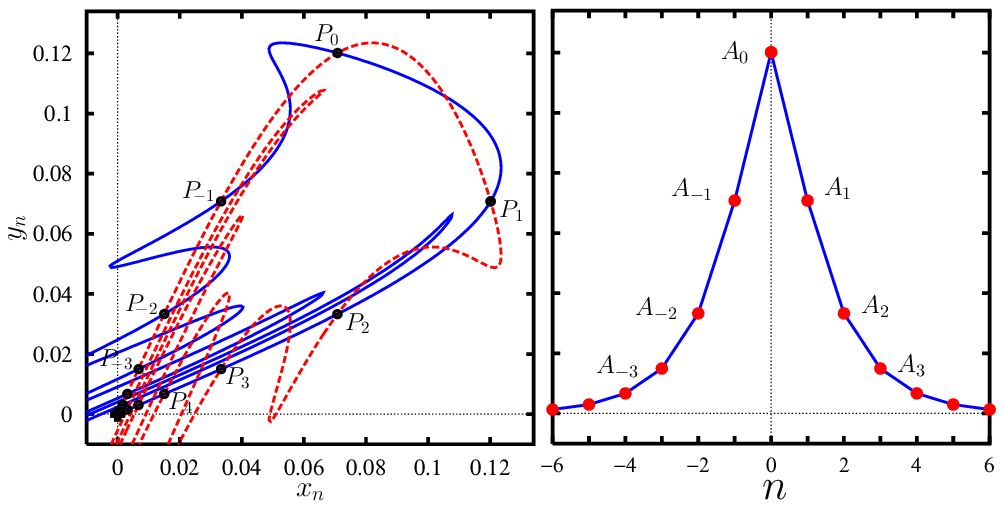}
} \smallskip
\centerline{~~
\includegraphics[width=8.25cm,angle=0]{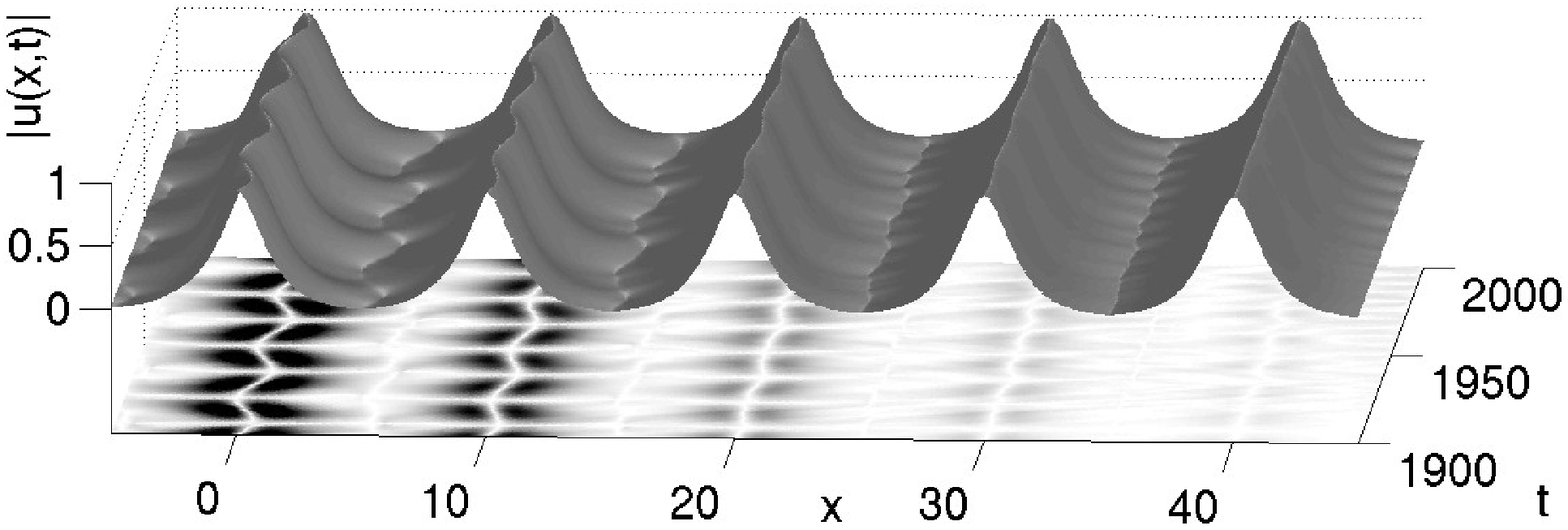}
} \caption{ Homoclinic connection of the origin (top-left panel)
[from Ref.\ \cite{promislow}],
giving rise to a spatially localized profile (top-right panel).
Bottom: the localized state in the original BEC model
(\protect\ref{nls3}) generated by the prescribed amplitude
configuration. The shaded base, depicting $\partial
|u|/\partial t$, highlights the areas in which the atomic 
density $|u(x,t)|$ varies the most.  Observe that the solution 
decays as one moves away from the center ($x=0$).  (As the 
solution is symmetric with respect to $x=0$, only its 
right-hand side is shown.)} \label{homoclinic}
\end{figure}

A typical intersection for the stable (solid curve) and
unstable (dashed curve) manifolds of the 2D map (\ref{2DM}) is
depicted in the top-left panel of Fig.\ \ref{homoclinic}. The
intersection points $P_{i}$ between the stable and unstable
manifolds then generate a localized configuration for the recurrence
relationship (\ref{recurrence}), as depicted in the top-right panel 
of Fig.\ \ref{homoclinic}. When this localized configuration is inserted 
into the original GP equation (\ref{nls3}), one obtains a spatially 
localized, multi-soliton state (depicted in the bottom panel of 
Fig.\ \ref{homoclinic}) by shifting each soliton from its equilibrium 
position by the prescribed amount \cite{promislow,ric2}.

It is important to note that generic localized initial configurations do not
give rise to long-lived, self-sustained ILMs. Nonetheless, the construction
described above is quite efficient in producing approximate initial
configurations that generate clean and robust localized states, such as the one
displayed in Fig.\ \ref{homoclinic}. The structural and dynamical stability
of these localized states is quite interesting. For example, the structural 
stability of the homoclinic tangle of the 2D map guarantees the existence of 
the ILM solution in the original model (\ref{nls3}), despite the employment of 
various approximations in the former's derivation. On the other hand, the 
dynamical stability of ILMs permits their observation even in the presence of 
strong perturbations (such as noise). Indeed, numerical experiments show that 
ILMs prevail even in the presence of large perturbations to the initial 
configuration or strong numerical noise, as discussed in 
Refs. \cite{promislow,ric2}.

Finally, it is worth mentioning that the dynamical reduction to
the 2D map (\ref{2DM}) can also be used to generate---in addition to the
localized states discussed above---a wide range of spatiotemporal 
structures by following the map's fixed points, periodic orbits, 
quasi-periodic orbits, and even chaotic orbits. Further, the
techniques described in this section can also be applied to chains of
bright solitons in which the deep OL is replaced by an
array of focused laser beams or impurities that tends to pin the
solitons and serve as a local effective attractive 
potential [see Eq.\ (\ref{xi})].

\section{Conclusions} \label{conc}

In this work, we have surveyed recent research on lattice
dynamics of Bose-Einstein condensates (BECs) in optical lattice (OL)
potentials. We discussed, in particular, the dynamics of BECs in a deep
OL that naturally discretizes the original spatially continuous
system into a collection of discrete interacting nodes (wells). The
ensuing nonlinear lattice features the trademark localized
solutions---solitons and intrinsic localized modes---that
have rendered the Fermi-Pasta-Ulam problem one of last century's most 
important contributions to nonlinear science. Specifically, using 
Wannier-function expansions, we reduced the dynamics of the 
Gross-Pitaevskii (GP) equation to a discrete nonlinear 
Schr\"{o}dinger (DNLS) equation, which was subsequently used to identify 
localized solutions (as well as a number of other interesting dynamical 
instabilities, such as those discussed in Ref. \cite{kef}) and reconstruct 
the nonlinear waves of the original GP equation.  Treating a 
BEC trapped in a strong OL as a collection of interacting solitons, 
we then used perturbative and/or variational techniques to reduce the 
original GP equation to a Toda-lattice equation with an effective 
on-site potential. This nonlinear lattice supports robust, 
exponentially localized oscillations in BEC soliton chains.

The crucial ingredient, amenable to the dynamical systems perspective
described in this manuscript, is the GP equation's nonlinear
self-interaction, which is induced as a mean-field representation of
two-particle interactions and is responsible for a plethora of
dynamical behaviors. The most remarkable of these, solitary waves
(solitons), lie at the heart of the paradigm that originated from the
seminal work of Fermi, Pasta, and Ulam.

There remain, moreover, a multitude of exciting directions for 
future research.  For example, the study of nonlinear lattice 
models in higher dimensions has yielded a number of interesting 
generalizations and more exotic solitary waves such as 
multi-dimensional solitons \cite{ijmpbrev} and discrete 
vortices \cite{borya1,borya2}.  A systematic analysis of their 
existence properties, dynamical characteristics (mobility and 
structural stability), thermodynamic properties, and physical 
relevance involves a number of interesting and subtle questions 
that may keep nonlinear scientists like us busy until we're ready to 
celebrate the 100th anniversary of the publication of the FPU problem.

\section*{ACKNOWLEDGMENTS}

We thank David Campbell for the invitation to write this article and
Predrag Cvitanovi\'c and Norm Zabusky for useful discussions during
the course of this work. P.G.K.\ gratefully acknowledges support from
NSF-DMS-0204585, from the Eppley Foundation for Research and from an
NSF-CAREER award. The work of B.A.M.\ was supported in a part by the
grant No. 8006/03 from the Israel Science Foundation. R.C.G.\ 
gratefully acknowledges support from an SDSU Grant-In-Aid.  MAP
acknowledges support provided by a VIGRE grant awarded to the School
of Mathematics at Georgia Tech.


\begin{thebibliography}{97}
\expandafter\ifx\csname natexlab\endcsname\relax\def\natexlab#1{#1}\fi
\expandafter\ifx\csname bibnamefont\endcsname\relax
  \def\bibnamefont#1{#1}\fi
\expandafter\ifx\csname bibfnamefont\endcsname\relax
  \def\bibfnamefont#1{#1}\fi
\expandafter\ifx\csname citenamefont\endcsname\relax
  \def\citenamefont#1{#1}\fi
\expandafter\ifx\csname url\endcsname\relax
  \def\url#1{\texttt{#1}}\fi
\expandafter\ifx\csname urlprefix\endcsname\relax\def\urlprefix{URL }\fi
\providecommand{\bibinfo}[2]{#2}
\providecommand{\eprint}[2][]{\url{#2}}

\bibitem[{\citenamefont{Weissert}(1997)}]{fpugen}
\bibinfo{author}{\bibfnamefont{T.~P.} \bibnamefont{Weissert}},
  \emph{\bibinfo{title}{The Genesis of Simulation in Dynamics: Pursuing the
  {F}ermi-{P}asta-{U}lam Problem}} (\bibinfo{publisher}{Springer-Verlag},
  \bibinfo{address}{New York, NY}, \bibinfo{year}{1997}).

\bibitem[{\citenamefont{Davis et~al.}(1995)\citenamefont{Davis, Mewes, Andrews,
  van Druten, Durfee, Kurn, and Ketterle}}]{becna}
\bibinfo{author}{\bibfnamefont{K.~B.} \bibnamefont{Davis}},
  \bibinfo{author}{\bibfnamefont{M.-O.} \bibnamefont{Mewes}},
  \bibinfo{author}{\bibfnamefont{M.~R.} \bibnamefont{Andrews}},
  \bibinfo{author}{\bibfnamefont{N.~J.} \bibnamefont{van Druten}},
  \bibinfo{author}{\bibfnamefont{D.~S.} \bibnamefont{Durfee}},
  \bibinfo{author}{\bibfnamefont{D.~M.} \bibnamefont{Kurn}}, \bibnamefont{and}
  \bibinfo{author}{\bibfnamefont{W.}~\bibnamefont{Ketterle}},
  \bibinfo{journal}{Phys.\ Rev.\ Lett.} \textbf{\bibinfo{volume}{75}},
  \bibinfo{pages}{3969} (\bibinfo{year}{1995}).

\bibitem[{\citenamefont{Anderson et~al.}(1995)\citenamefont{Anderson, Ensher,
  Matthews, Wieman, and Cornell}}]{becrub}
\bibinfo{author}{\bibfnamefont{M.~H.} \bibnamefont{Anderson}},
  \bibinfo{author}{\bibfnamefont{J.~R.} \bibnamefont{Ensher}},
  \bibinfo{author}{\bibfnamefont{M.~R.} \bibnamefont{Matthews}},
  \bibinfo{author}{\bibfnamefont{C.~E.} \bibnamefont{Wieman}},
  \bibnamefont{and} \bibinfo{author}{\bibfnamefont{E.~A.}
  \bibnamefont{Cornell}}, \bibinfo{journal}{Science}
  \textbf{\bibinfo{volume}{269}}, \bibinfo{pages}{198} (\bibinfo{year}{1995}).

\bibitem[{\citenamefont{Zabusky}(1984)}]{norm}
\bibinfo{author}{\bibfnamefont{N.~J.} \bibnamefont{Zabusky}},
  \bibinfo{journal}{Physics Today} pp. \bibinfo{pages}{2--12}
  (\bibinfo{year}{1984}).

\bibitem[{\citenamefont{Friesecke and Pego}(1999)}]{friep1}
\bibinfo{author}{\bibfnamefont{G.}~\bibnamefont{Friesecke}} \bibnamefont{and}
  \bibinfo{author}{\bibfnamefont{R.~L.} \bibnamefont{Pego}},
  \bibinfo{journal}{Nonlinearity} \textbf{\bibinfo{volume}{12}},
  \bibinfo{pages}{1601} (\bibinfo{year}{1999}).

\bibitem[{\citenamefont{Friesecke and Pego}(2002)}]{friep2}
\bibinfo{author}{\bibfnamefont{G.}~\bibnamefont{Friesecke}} \bibnamefont{and}
  \bibinfo{author}{\bibfnamefont{R.~L.} \bibnamefont{Pego}},
  \bibinfo{journal}{Nonlinearity} \textbf{\bibinfo{volume}{15}},
  \bibinfo{pages}{1343} (\bibinfo{year}{2002}).

\bibitem[{\citenamefont{Friesecke and Pego}(2004{\natexlab{a}})}]{friep3}
\bibinfo{author}{\bibfnamefont{G.}~\bibnamefont{Friesecke}} \bibnamefont{and}
  \bibinfo{author}{\bibfnamefont{R.~L.} \bibnamefont{Pego}},
  \bibinfo{journal}{Nonlinearity} \textbf{\bibinfo{volume}{17}},
  \bibinfo{pages}{207} (\bibinfo{year}{2004}{\natexlab{a}}).

\bibitem[{\citenamefont{Friesecke and Pego}(2004{\natexlab{b}})}]{friep4}
\bibinfo{author}{\bibfnamefont{G.}~\bibnamefont{Friesecke}} \bibnamefont{and}
  \bibinfo{author}{\bibfnamefont{R.~L.} \bibnamefont{Pego}},
  \bibinfo{journal}{Nonlinearity} \textbf{\bibinfo{volume}{17}},
  \bibinfo{pages}{229} (\bibinfo{year}{2004}{\natexlab{b}}).

\bibitem[{\citenamefont{Lepri et~al.}(2003)\citenamefont{Lepri, Livi, and
  Politi}}]{Lepri}
\bibinfo{author}{\bibfnamefont{S.}~\bibnamefont{Lepri}},
  \bibinfo{author}{\bibfnamefont{R.}~\bibnamefont{Livi}}, \bibnamefont{and}
  \bibinfo{author}{\bibfnamefont{A.}~\bibnamefont{Politi}},
  \bibinfo{journal}{Physics Reports} \textbf{\bibinfo{volume}{377}},
  \bibinfo{pages}{1} (\bibinfo{year}{2003}).

\bibitem[{\citenamefont{Fermi et~al.}(1955)\citenamefont{Fermi, Pasta, and
  Ulam}}]{fpu55}
\bibinfo{author}{\bibfnamefont{E.}~\bibnamefont{Fermi}},
  \bibinfo{author}{\bibfnamefont{J.~R.} \bibnamefont{Pasta}}, \bibnamefont{and}
  \bibinfo{author}{\bibfnamefont{S.}~\bibnamefont{Ulam}}, \bibinfo{type}{Tech.
  Rep.} \bibinfo{number}{Report LA-1940}, \bibinfo{institution}{Los Alamos}
  (\bibinfo{year}{1955}).

\bibitem[{\citenamefont{Lichtenberg and Lieberman}(1992)}]{lich}
\bibinfo{author}{\bibfnamefont{A.~J.} \bibnamefont{Lichtenberg}}
  \bibnamefont{and} \bibinfo{author}{\bibfnamefont{M.~A.}
  \bibnamefont{Lieberman}}, \emph{\bibinfo{title}{Regular and Chaotic
  Dynamics}}, no.~\bibinfo{number}{38} in \bibinfo{series}{Applied Mathematical
  Sciences} (\bibinfo{publisher}{Springer-Verlag}, \bibinfo{address}{New York,
  NY}, \bibinfo{year}{1992}), \bibinfo{edition}{2nd} ed.

\bibitem[{\citenamefont{Ablowitz and Clarkson}(1991)}]{ablo}
\bibinfo{author}{\bibfnamefont{M.~J.} \bibnamefont{Ablowitz}} \bibnamefont{and}
  \bibinfo{author}{\bibfnamefont{P.~A.} \bibnamefont{Clarkson}},
  \emph{\bibinfo{title}{Solitons, Nonlinear Evolution Equations and Inverse
  Scattering}}, no. \bibinfo{number}{149} in \bibinfo{series}{London
  Mathematical Society Lecture Notes Series} (\bibinfo{publisher}{Cambridge
  University Press}, \bibinfo{address}{Cambridge, U.K.}, \bibinfo{year}{1991}).

\bibitem[{\citenamefont{Zabusky and Kruskal}(1965)}]{zab}
\bibinfo{author}{\bibfnamefont{N.~J.} \bibnamefont{Zabusky}} \bibnamefont{and}
  \bibinfo{author}{\bibfnamefont{M.~D.} \bibnamefont{Kruskal}},
  \bibinfo{journal}{Phys.\ Rev.\ Lett.} \textbf{\bibinfo{volume}{15}},
  \bibinfo{pages}{240} (\bibinfo{year}{1965}).

\bibitem[{\citenamefont{Dodd et~al.}(1982)\citenamefont{Dodd, Eilbeck, Gibbon,
  and Morris}}]{JCE}
\bibinfo{author}{\bibfnamefont{R.}~\bibnamefont{Dodd}},
  \bibinfo{author}{\bibfnamefont{J.}~\bibnamefont{Eilbeck}},
  \bibinfo{author}{\bibfnamefont{J.}~\bibnamefont{Gibbon}}, \bibnamefont{and}
  \bibinfo{author}{\bibfnamefont{H.}~\bibnamefont{Morris}},
  \emph{\bibinfo{title}{Solitons and Nonlinear Wave Equations}}
  (\bibinfo{publisher}{London: Academic Press}, \bibinfo{address}{London, UK},
  \bibinfo{year}{1982}).

\bibitem[{\citenamefont{Drazin and Johnson}(1989)}]{Drazin}
\bibinfo{author}{\bibfnamefont{P.}~\bibnamefont{Drazin}} \bibnamefont{and}
  \bibinfo{author}{\bibfnamefont{R.}~\bibnamefont{Johnson}},
  \emph{\bibinfo{title}{Solitons: An Introduction}}
  (\bibinfo{publisher}{Cambridge University Press},
  \bibinfo{address}{Cambridge, UK}, \bibinfo{year}{1989}).

\bibitem[{\citenamefont{Whitham}(1974)}]{whitham}
\bibinfo{author}{\bibfnamefont{G.~B.} \bibnamefont{Whitham}},
  \emph{\bibinfo{title}{Linear and Nonlinear Waves}}, Pure and Applied
  Mathematics (\bibinfo{publisher}{Wiley-Interscience}, \bibinfo{address}{New
  York, NY}, \bibinfo{year}{1974}).

\bibitem[{\citenamefont{Ablowitz and Segur}(1981)}]{ist}
\bibinfo{author}{\bibfnamefont{M.~J.} \bibnamefont{Ablowitz}} \bibnamefont{and}
  \bibinfo{author}{\bibfnamefont{H.}~\bibnamefont{Segur}},
  \emph{\bibinfo{title}{Solitons and the Inverse Scattering Transform}}, SIAM
  Studies in Applied Mathematics (\bibinfo{publisher}{Society for Industrial
  and Applied Mathematics}, \bibinfo{address}{Philadelphia, Pennsylvania},
  \bibinfo{year}{1981}).

\bibitem[{\citenamefont{Pethick and Smith}(2002)}]{pethick}
\bibinfo{author}{\bibfnamefont{C.~J.} \bibnamefont{Pethick}} \bibnamefont{and}
  \bibinfo{author}{\bibfnamefont{H.}~\bibnamefont{Smith}},
  \emph{\bibinfo{title}{Bose-Einstein Condensation in Dilute Gases}}
  (\bibinfo{publisher}{Cambridge University Press},
  \bibinfo{address}{Cambridge, United Kingdom}, \bibinfo{year}{2002}).

\bibitem[{\citenamefont{Dalfovo et~al.}(1999)\citenamefont{Dalfovo, Giorgini,
  Pitaevskii, and Stringari}}]{stringari}
\bibinfo{author}{\bibfnamefont{F.}~\bibnamefont{Dalfovo}},
  \bibinfo{author}{\bibfnamefont{S.}~\bibnamefont{Giorgini}},
  \bibinfo{author}{\bibfnamefont{L.~P.} \bibnamefont{Pitaevskii}},
  \bibnamefont{and}
  \bibinfo{author}{\bibfnamefont{S.}~\bibnamefont{Stringari}},
  \bibinfo{journal}{Rev.\ Mod.\ Phys.} \textbf{\bibinfo{volume}{71}},
  \bibinfo{pages}{463} (\bibinfo{year}{1999}).

\bibitem[{\citenamefont{Alfimov
  et~al.}(2002{\natexlab{a}})\citenamefont{Alfimov, Kevrekidis, Konotop, and
  Salerno}}]{alf}
\bibinfo{author}{\bibfnamefont{G.~L.} \bibnamefont{Alfimov}},
  \bibinfo{author}{\bibfnamefont{P.~G.} \bibnamefont{Kevrekidis}},
  \bibinfo{author}{\bibfnamefont{V.~V.} \bibnamefont{Konotop}},
  \bibnamefont{and} \bibinfo{author}{\bibfnamefont{M.}~\bibnamefont{Salerno}},
  \bibinfo{journal}{Phys.\ Rev.\ E} \textbf{\bibinfo{volume}{66}}
  (\bibinfo{year}{2002}{\natexlab{a}}).

\bibitem[{\citenamefont{Izrailev and Chirikov}(1966)}]{iz}
\bibinfo{author}{\bibfnamefont{F.~M.} \bibnamefont{Izrailev}} \bibnamefont{and}
  \bibinfo{author}{\bibfnamefont{B.~V.} \bibnamefont{Chirikov}},
  \bibinfo{journal}{Soviet Physics Doklady} \textbf{\bibinfo{volume}{11}},
  \bibinfo{pages}{30} (\bibinfo{year}{1966}).

\bibitem[{\citenamefont{Driscoll and O'Neil}(1976)}]{Driscoll:76}
\bibinfo{author}{\bibfnamefont{C.~F.} \bibnamefont{Driscoll}} \bibnamefont{and}
  \bibinfo{author}{\bibfnamefont{T.~M.} \bibnamefont{O'Neil}},
  \bibinfo{journal}{Phys.\ Rev.\ Lett.} \textbf{\bibinfo{volume}{37}},
  \bibinfo{pages}{69} (\bibinfo{year}{1976}).

\bibitem[{\citenamefont{Kruskal}(1975)}]{krusk}
\bibinfo{author}{\bibfnamefont{M.}~\bibnamefont{Kruskal}}, in
  \emph{\bibinfo{booktitle}{Dynamical Systems Theory and Applications}}, edited
  by \bibinfo{editor}{\bibfnamefont{J.}~\bibnamefont{Moser}}
  (\bibinfo{publisher}{Springer-Verlag}, \bibinfo{address}{New York, NY},
  \bibinfo{year}{1975}), vol.~\bibinfo{volume}{38} of
  \emph{\bibinfo{series}{Lecture Notes in Physics}}, pp.
  \bibinfo{pages}{310--354}, \bibinfo{note}{Seattle 1974}.

\bibitem[{\citenamefont{Newell}(1974)}]{nwm}
\bibinfo{editor}{\bibfnamefont{A.~C.} \bibnamefont{Newell}}, ed.,
  \emph{\bibinfo{title}{Nonlinear Wave Motion}}, vol.~\bibinfo{volume}{15} of
  \emph{\bibinfo{series}{Lectures in Applied Mathematics}}
  (\bibinfo{publisher}{American Mathematical Society},
  \bibinfo{address}{Providence, Rhode Island}, \bibinfo{year}{1974}).

\bibitem[{\citenamefont{Rink}(2002)}]{rink1}
\bibinfo{author}{\bibfnamefont{B.}~\bibnamefont{Rink}}, \bibinfo{journal}{J.\
  Nonlin.\ Sci.} \textbf{\bibinfo{volume}{12}}, \bibinfo{pages}{479}
  (\bibinfo{year}{2002}).

\bibitem[{\citenamefont{Rink}(2003)}]{rink2}
\bibinfo{author}{\bibfnamefont{B.}~\bibnamefont{Rink}},
  \bibinfo{journal}{Physica D} \textbf{\bibinfo{volume}{175}},
  \bibinfo{pages}{31} (\bibinfo{year}{2003}).

\bibitem[{\citenamefont{Rink}(2001)}]{rink3}
\bibinfo{author}{\bibfnamefont{B.}~\bibnamefont{Rink}},
  \bibinfo{journal}{Comm.\ Math.\ Phys.} \textbf{\bibinfo{volume}{218}},
  \bibinfo{pages}{665} (\bibinfo{year}{2001}).
  
\bibitem[{\citenamefont{Biello et~al.}(2002)\citenamefont{Biello,
      Kramer, and Lvov}}]{Biello}
  \bibinfo{author}\bibinfo{author}{\bibfnamefont{P.~R.}
    \bibnamefont{Kramer}}, {\bibfnamefont{J.~A.}
    \bibnamefont{Biello}}, \bibnamefont{and}
  \bibinfo{author}{\bibfnamefont{Y.}~\bibnamefont{Lvov}}, in
  \bibinfo{journal}{Proceedings of the 4th International Conference on
    Dynamical Systems and Differential Equations}
  (\bibinfo{year}{2002}), \bibinfo{note}{nlin.CD/0210007}.

\bibitem[{\citenamefont{Gerdjikov et~al.}(1996)\citenamefont{Gerdjikov, Kaup,
  Uzunov, and Evstatiev}}]{kaup}
\bibinfo{author}{\bibfnamefont{V.~S.} \bibnamefont{Gerdjikov}},
  \bibinfo{author}{\bibfnamefont{D.~J.} \bibnamefont{Kaup}},
  \bibinfo{author}{\bibfnamefont{I.~M.} \bibnamefont{Uzunov}},
  \bibnamefont{and} \bibinfo{author}{\bibfnamefont{E.~G.}
  \bibnamefont{Evstatiev}}, \bibinfo{journal}{Phys.\ Rev.\ Lett.}
  \textbf{\bibinfo{volume}{77}}, \bibinfo{pages}{3943} (\bibinfo{year}{1996}).

\bibitem[{\citenamefont{Berman and Kolovskii}(1984)}]{berkol}
\bibinfo{author}{\bibfnamefont{G.~P.} \bibnamefont{Berman}} \bibnamefont{and}
  \bibinfo{author}{\bibfnamefont{A.~R.} \bibnamefont{Kolovskii}},
  \bibinfo{journal}{Soviet Physics JETP} \textbf{\bibinfo{volume}{60}},
  \bibinfo{pages}{1116} (\bibinfo{year}{1984}).

\bibitem[{\citenamefont{Ketterle}(1999)}]{ketter}
\bibinfo{author}{\bibfnamefont{W.}~\bibnamefont{Ketterle}},
  \bibinfo{journal}{Physics Today} \textbf{\bibinfo{volume}{52}},
  \bibinfo{pages}{30} (\bibinfo{year}{1999}).

\bibitem[{\citenamefont{Burnett et~al.}(1999)\citenamefont{Burnett, Edwards,
  and Clark}}]{edwards}
\bibinfo{author}{\bibfnamefont{K.}~\bibnamefont{Burnett}},
  \bibinfo{author}{\bibfnamefont{M.}~\bibnamefont{Edwards}}, \bibnamefont{and}
  \bibinfo{author}{\bibfnamefont{C.~W.} \bibnamefont{Clark}},
  \bibinfo{journal}{Physics Today} \textbf{\bibinfo{volume}{52}},
  \bibinfo{pages}{37} (\bibinfo{year}{1999}).

\bibitem[{\citenamefont{K\"ohler}(2002)}]{kohler}
\bibinfo{author}{\bibfnamefont{T.}~\bibnamefont{K\"ohler}},
  \bibinfo{journal}{Phys.\ Rev.\ Lett.} \textbf{\bibinfo{volume}{89}},
  \bibinfo{pages}{210404} (\bibinfo{year}{2002}).

\bibitem[{\citenamefont{Baizakov et~al.}(2002)\citenamefont{Baizakov, Konotop,
  and Salerno}}]{baiz}
\bibinfo{author}{\bibfnamefont{B.~B.} \bibnamefont{Baizakov}},
  \bibinfo{author}{\bibfnamefont{V.~V.} \bibnamefont{Konotop}},
  \bibnamefont{and} \bibinfo{author}{\bibfnamefont{M.}~\bibnamefont{Salerno}},
  \bibinfo{journal}{J.\ Phys.\ B: Atom.\ Mol.\ Phys.}
  \textbf{\bibinfo{volume}{35}}, \bibinfo{pages}{5105} (\bibinfo{year}{2002}).

\bibitem[{\citenamefont{Salasnich
  et~al.}(2002{\natexlab{a}})\citenamefont{Salasnich, Parola, and
  Reatto}}]{salasnich}
\bibinfo{author}{\bibfnamefont{L.}~\bibnamefont{Salasnich}},
  \bibinfo{author}{\bibfnamefont{A.}~\bibnamefont{Parola}}, \bibnamefont{and}
  \bibinfo{author}{\bibfnamefont{L.}~\bibnamefont{Reatto}},
  \bibinfo{journal}{J.\ Phys.\ B: Atom.\ Mol.\ Phys.}
  \textbf{\bibinfo{volume}{35}}, \bibinfo{pages}{3205}
  (\bibinfo{year}{2002}{\natexlab{a}}).

\bibitem[{\citenamefont{P\'erez-Garc\'{\i}a
  et~al.}(1998)\citenamefont{P\'erez-Garc\'{\i}a, Michinel, and
  Herrero}}]{Perez-Garcia:98}
\bibinfo{author}{\bibfnamefont{V.}~\bibnamefont{P\'erez-Garc\'{\i}a}},
  \bibinfo{author}{\bibfnamefont{H.}~\bibnamefont{Michinel}}, \bibnamefont{and}
  \bibinfo{author}{\bibfnamefont{H.}~\bibnamefont{Herrero}},
  \bibinfo{journal}{Phys.\ Rev.\ A} \textbf{\bibinfo{volume}{57}},
  \bibinfo{pages}{3837} (\bibinfo{year}{1998}).

\bibitem[{\citenamefont{Salasnich
  et~al.}(2002{\natexlab{b}})\citenamefont{Salasnich, Parola, and
  Reatto}}]{Salasnich:02a}
\bibinfo{author}{\bibfnamefont{L.}~\bibnamefont{Salasnich}},
  \bibinfo{author}{\bibfnamefont{A.}~\bibnamefont{Parola}}, \bibnamefont{and}
  \bibinfo{author}{\bibfnamefont{L.}~\bibnamefont{Reatto}},
  \bibinfo{journal}{Phys.\ Rev.\ A} \textbf{\bibinfo{volume}{66}},
  \bibinfo{pages}{043603} (\bibinfo{year}{2002}{\natexlab{b}}).

\bibitem[{\citenamefont{Band et~al.}(2003)\citenamefont{Band, Towers, and
  Malomed}}]{Boris:03}
\bibinfo{author}{\bibfnamefont{Y.~B.} \bibnamefont{Band}},
  \bibinfo{author}{\bibfnamefont{I.}~\bibnamefont{Towers}}, \bibnamefont{and}
  \bibinfo{author}{\bibfnamefont{B.~A.} \bibnamefont{Malomed}},
  \bibinfo{journal}{Phys.\ Rev.\ A} \textbf{\bibinfo{volume}{67}},
  \bibinfo{pages}{023602} (\bibinfo{year}{2003}).

\bibitem[{\citenamefont{Lieb et~al.}(2003)\citenamefont{Lieb, Seiringer, and
  Yngvason}}]{Lieb:03}
\bibinfo{author}{\bibfnamefont{E.~H.} \bibnamefont{Lieb}},
  \bibinfo{author}{\bibfnamefont{R.}~\bibnamefont{Seiringer}},
  \bibnamefont{and} \bibinfo{author}{\bibfnamefont{J.}~\bibnamefont{Yngvason}},
  \bibinfo{journal}{Phys.\ Rev.\ Lett.} \textbf{\bibinfo{volume}{91}},
  \bibinfo{pages}{150401} (\bibinfo{year}{2003}).

\bibitem[{\citenamefont{Gross}(1961)}]{Gross:61}
\bibinfo{author}{\bibfnamefont{E.~P.} \bibnamefont{Gross}},
  \bibinfo{journal}{Nuovo Cim.} \textbf{\bibinfo{volume}{20}},
  \bibinfo{pages}{454} (\bibinfo{year}{1961}).

\bibitem[{\citenamefont{Pitaevskii}(1961)}]{Pitaevskii:61}
\bibinfo{author}{\bibfnamefont{L.~P.} \bibnamefont{Pitaevskii}},
  \bibinfo{journal}{Sov.\ Phys.\ JETP} \textbf{\bibinfo{volume}{13}},
  \bibinfo{pages}{451} (\bibinfo{year}{1961}).

\bibitem[{\citenamefont{Donley et~al.}(2001)\citenamefont{Donley, Claussen,
  Cornish, Roberts, Cornell, and Weiman}}]{fesh}
\bibinfo{author}{\bibfnamefont{E.~A.} \bibnamefont{Donley}},
  \bibinfo{author}{\bibfnamefont{N.~R.} \bibnamefont{Claussen}},
  \bibinfo{author}{\bibfnamefont{S.~L.} \bibnamefont{Cornish}},
  \bibinfo{author}{\bibfnamefont{J.~L.} \bibnamefont{Roberts}},
  \bibinfo{author}{\bibfnamefont{E.~A.} \bibnamefont{Cornell}},
  \bibnamefont{and} \bibinfo{author}{\bibfnamefont{C.~E.}
  \bibnamefont{Weiman}}, \bibinfo{journal}{Nature}
  \textbf{\bibinfo{volume}{412}}, \bibinfo{pages}{295} (\bibinfo{year}{2001}).

\bibitem[{\citenamefont{Kevrekidis
  et~al.}(2003{\natexlab{a}})\citenamefont{Kevrekidis, Theocharis,
  Frantzeskakis, and Malomed}}]{FRM}
\bibinfo{author}{\bibfnamefont{P.~G.} \bibnamefont{Kevrekidis}},
  \bibinfo{author}{\bibfnamefont{G.}~\bibnamefont{Theocharis}},
  \bibinfo{author}{\bibfnamefont{D.~J.} \bibnamefont{Frantzeskakis}},
  \bibnamefont{and} \bibinfo{author}{\bibfnamefont{B.~A.}
  \bibnamefont{Malomed}}, \bibinfo{journal}{Phys.\ Rev.\ Lett.}
  \textbf{\bibinfo{volume}{90}}, \bibinfo{pages}{230401}
  (\bibinfo{year}{2003}{\natexlab{a}}).

\bibitem[{\citenamefont{Anderson and
  Kasevich}(1998{\natexlab{a}})}]{Anderson:98}
\bibinfo{author}{\bibfnamefont{B.~P.} \bibnamefont{Anderson}} \bibnamefont{and}
  \bibinfo{author}{\bibfnamefont{M.~A.} \bibnamefont{Kasevich}},
  \bibinfo{journal}{Science} \textbf{\bibinfo{volume}{282}},
  \bibinfo{pages}{1686} (\bibinfo{year}{1998}{\natexlab{a}}).

\bibitem[{\citenamefont{Jona-Lasinio et~al.}(2003)\citenamefont{Jona-Lasinio,
  Morsch, Cristiani, Malossi, M\"uller, Courtade, Anderlini, and Arimondo}}]{g2b}
\bibinfo{author}{\bibfnamefont{M.}~\bibnamefont{Jona-Lasinio}},
  \bibinfo{author}{\bibfnamefont{O.}~\bibnamefont{Morsch}},
  \bibinfo{author}{\bibfnamefont{M.}~\bibnamefont{Cristiani}},
  \bibinfo{author}{\bibfnamefont{N.}~\bibnamefont{Malossi}},
  \bibinfo{author}{\bibfnamefont{J.~H.} \bibnamefont{M\"uller}},
  \bibinfo{author}{\bibfnamefont{E.}~\bibnamefont{Courtade}},
  \bibinfo{author}{\bibfnamefont{M.}~\bibnamefont{Anderlini}},
  \bibnamefont{and} \bibinfo{author}{\bibfnamefont{E.}~\bibnamefont{Arimondo}},
  \bibinfo{journal}{Phys.\ Rev.\ Lett.} \textbf{\bibinfo{volume}{91}},
  \bibinfo{pages}{230406} (\bibinfo{year}{2003}).

\bibitem[{\citenamefont{Konotop et~al.}()\citenamefont{Konotop, Kevrekidis, and
  Salerno}}]{g2c}
\bibinfo{author}{\bibfnamefont{V.~V.} \bibnamefont{Konotop}},
  \bibinfo{author}{\bibfnamefont{P.~G.} \bibnamefont{Kevrekidis}},
  \bibnamefont{and} \bibinfo{author}{\bibfnamefont{M.}~\bibnamefont{Salerno}},
  \bibinfo{note}{cond-mat/0404608}.

\bibitem[{\citenamefont{Orzel et~al.}(2001{\natexlab{a}})\citenamefont{Orzel,
  Tuchman, Fenselau, Yasuda, and Kasevich}}]{g3}
\bibinfo{author}{\bibfnamefont{C.}~\bibnamefont{Orzel}},
  \bibinfo{author}{\bibfnamefont{A.~K.} \bibnamefont{Tuchman}},
  \bibinfo{author}{\bibfnamefont{M.~L.} \bibnamefont{Fenselau}},
  \bibinfo{author}{\bibfnamefont{M.}~\bibnamefont{Yasuda}}, \bibnamefont{and}
  \bibinfo{author}{\bibfnamefont{M.~A.} \bibnamefont{Kasevich}},
  \bibinfo{journal}{Science} \textbf{\bibinfo{volume}{291}},
  \bibinfo{pages}{2386} (\bibinfo{year}{2001}{\natexlab{a}}).

\bibitem[{\citenamefont{Morsch et~al.}(2001{\natexlab{a}})\citenamefont{Morsch,
  M\"uller, Cristiani, Ciampini, and Arimondo}}]{g4}
\bibinfo{author}{\bibfnamefont{O.}~\bibnamefont{Morsch}},
  \bibinfo{author}{\bibfnamefont{J.~H.} \bibnamefont{M\"uller}},
  \bibinfo{author}{\bibfnamefont{M.}~\bibnamefont{Cristiani}},
  \bibinfo{author}{\bibfnamefont{D.}~\bibnamefont{Ciampini}}, \bibnamefont{and}
  \bibinfo{author}{\bibfnamefont{E.}~\bibnamefont{Arimondo}},
  \bibinfo{journal}{Phys.\ Rev.\ Lett.} \textbf{\bibinfo{volume}{87}},
  \bibinfo{pages}{140402} (\bibinfo{year}{2001}{\natexlab{a}}).

\bibitem[{\citenamefont{Cataliotti et~al.}(2001)\citenamefont{Cataliotti,
  Burger, Fort, Maddaloni, Minardi, Trombettoni, Smerzi, and
  Inguscio}}]{Smerzi:01}
\bibinfo{author}{\bibfnamefont{F.~S.} \bibnamefont{Cataliotti}},
  \bibinfo{author}{\bibfnamefont{S.}~\bibnamefont{Burger}},
  \bibinfo{author}{\bibfnamefont{C.}~\bibnamefont{Fort}},
  \bibinfo{author}{\bibfnamefont{P.}~\bibnamefont{Maddaloni}},
  \bibinfo{author}{\bibfnamefont{F.}~\bibnamefont{Minardi}},
  \bibinfo{author}{\bibfnamefont{A.}~\bibnamefont{Trombettoni}},
  \bibinfo{author}{\bibfnamefont{A.}~\bibnamefont{Smerzi}}, \bibnamefont{and}
  \bibinfo{author}{\bibfnamefont{M.}~\bibnamefont{Inguscio}},
  \bibinfo{journal}{Science} \textbf{\bibinfo{volume}{293}},
  \bibinfo{pages}{843} (\bibinfo{year}{2001}).

\bibitem[{\citenamefont{Greiner et~al.}(2001)\citenamefont{Greiner, Bloch,
  Mandel, H\"ansch, and Esslinger}}]{g6}
\bibinfo{author}{\bibfnamefont{M.}~\bibnamefont{Greiner}},
  \bibinfo{author}{\bibfnamefont{I.}~\bibnamefont{Bloch}},
  \bibinfo{author}{\bibfnamefont{O.}~\bibnamefont{Mandel}},
  \bibinfo{author}{\bibfnamefont{T.~W.} \bibnamefont{H\"ansch}},
  \bibnamefont{and}
  \bibinfo{author}{\bibfnamefont{T.}~\bibnamefont{Esslinger}},
  \bibinfo{journal}{Phys.\ Rev.\ Lett.} \textbf{\bibinfo{volume}{87}},
  \bibinfo{pages}{160405} (\bibinfo{year}{2001}).

\bibitem[{\citenamefont{Greiner
  et~al.}(2002{\natexlab{a}})\citenamefont{Greiner, Mandel, Esslinger,
  H\"ansch, and Bloch}}]{g7}
\bibinfo{author}{\bibfnamefont{M.}~\bibnamefont{Greiner}},
  \bibinfo{author}{\bibfnamefont{O.}~\bibnamefont{Mandel}},
  \bibinfo{author}{\bibfnamefont{T.}~\bibnamefont{Esslinger}},
  \bibinfo{author}{\bibfnamefont{T.~W.} \bibnamefont{H\"ansch}},
  \bibnamefont{and} \bibinfo{author}{\bibfnamefont{I.}~\bibnamefont{Bloch}},
  \bibinfo{journal}{Nature} \textbf{\bibinfo{volume}{415}}, \bibinfo{pages}{39}
  (\bibinfo{year}{2002}{\natexlab{a}}).

\bibitem[{\citenamefont{Anderson and Kasevich}(1998{\natexlab{b}})}]{anderson}
\bibinfo{author}{\bibfnamefont{B.~P.} \bibnamefont{Anderson}} \bibnamefont{and}
  \bibinfo{author}{\bibfnamefont{M.~A.} \bibnamefont{Kasevich}},
  \bibinfo{journal}{Science} \textbf{\bibinfo{volume}{282}},
  \bibinfo{pages}{1686} (\bibinfo{year}{1998}{\natexlab{b}}).

\bibitem[{\citenamefont{Greiner
  et~al.}(2002{\natexlab{b}})\citenamefont{Greiner, Mandel, Esslinger,
  H\"ansch, and Bloch}}]{mott}
\bibinfo{author}{\bibfnamefont{M.}~\bibnamefont{Greiner}},
  \bibinfo{author}{\bibfnamefont{O.}~\bibnamefont{Mandel}},
  \bibinfo{author}{\bibfnamefont{T.}~\bibnamefont{Esslinger}},
  \bibinfo{author}{\bibfnamefont{T.}~\bibnamefont{H\"ansch}}, \bibnamefont{and}
  \bibinfo{author}{\bibfnamefont{I.}~\bibnamefont{Bloch}},
  \bibinfo{journal}{Nature} \textbf{\bibinfo{volume}{415}}
  (\bibinfo{year}{2002}{\natexlab{b}}).

\bibitem[{\citenamefont{Burger et~al.}(2001)\citenamefont{Burger, Cataliotti,
  Fort, Minardi, and Inguscio}}]{lattice}
\bibinfo{author}{\bibfnamefont{S.}~\bibnamefont{Burger}},
  \bibinfo{author}{\bibfnamefont{F.~S.} \bibnamefont{Cataliotti}},
  \bibinfo{author}{\bibfnamefont{C.}~\bibnamefont{Fort}},
  \bibinfo{author}{\bibfnamefont{F.}~\bibnamefont{Minardi}}, \bibnamefont{and}
  \bibinfo{author}{\bibfnamefont{M.}~\bibnamefont{Inguscio}},
  \bibinfo{journal}{Phys.\ Rev.\ Lett.} \textbf{\bibinfo{volume}{86}},
  \bibinfo{pages}{4447} (\bibinfo{year}{2001}).

\bibitem[{\citenamefont{Morsch et~al.}(2001{\natexlab{b}})\citenamefont{Morsch,
  M\"uller, Christiani, Ciampini, and Arimondo}}]{morsch}
\bibinfo{author}{\bibfnamefont{O.}~\bibnamefont{Morsch}},
  \bibinfo{author}{\bibfnamefont{J.~H.} \bibnamefont{M\"uller}},
  \bibinfo{author}{\bibfnamefont{M.}~\bibnamefont{Christiani}},
  \bibinfo{author}{\bibfnamefont{D.}~\bibnamefont{Ciampini}}, \bibnamefont{and}
  \bibinfo{author}{\bibfnamefont{E.}~\bibnamefont{Arimondo}},
  \bibinfo{journal}{Phys.\ Rev.\ Lett.} \textbf{\bibinfo{volume}{87}}
  (\bibinfo{year}{2001}{\natexlab{b}}).

\bibitem[{\citenamefont{Choi and Niu}(1999)}]{space2}
\bibinfo{author}{\bibfnamefont{D.-I.} \bibnamefont{Choi}} \bibnamefont{and}
  \bibinfo{author}{\bibfnamefont{Q.}~\bibnamefont{Niu}},
  \bibinfo{journal}{Phys.\ Rev.\ Lett.} \textbf{\bibinfo{volume}{82}},
  \bibinfo{pages}{2022} (\bibinfo{year}{1999}).

\bibitem[{\citenamefont{Jaksch et~al.}(1998)\citenamefont{Jaksch, Bruder,
  Cirac, Gardiner, and Zoller}}]{tight}
\bibinfo{author}{\bibfnamefont{D.}~\bibnamefont{Jaksch}},
  \bibinfo{author}{\bibfnamefont{C.}~\bibnamefont{Bruder}},
  \bibinfo{author}{\bibfnamefont{J.~I.} \bibnamefont{Cirac}},
  \bibinfo{author}{\bibfnamefont{C.~W.} \bibnamefont{Gardiner}},
  \bibnamefont{and} \bibinfo{author}{\bibfnamefont{P.}~\bibnamefont{Zoller}},
  \bibinfo{journal}{Phys.\ Rev.\ Lett.} \textbf{\bibinfo{volume}{81}},
  \bibinfo{pages}{3108} (\bibinfo{year}{1998}).

\bibitem[{\citenamefont{Peil et~al.}(2003)\citenamefont{Peil, Porto, {Laburthe
  Tolra}, Obrecht, King, Subbotin, Rolston, and Phillips}}]{quasibec}
\bibinfo{author}{\bibfnamefont{S.}~\bibnamefont{Peil}},
  \bibinfo{author}{\bibfnamefont{J.~V.} \bibnamefont{Porto}},
  \bibinfo{author}{\bibfnamefont{B.}~\bibnamefont{{Laburthe Tolra}}},
  \bibinfo{author}{\bibfnamefont{J.~M.} \bibnamefont{Obrecht}},
  \bibinfo{author}{\bibfnamefont{B.~E.} \bibnamefont{King}},
  \bibinfo{author}{\bibfnamefont{M.}~\bibnamefont{Subbotin}},
  \bibinfo{author}{\bibfnamefont{S.~L.} \bibnamefont{Rolston}},
  \bibnamefont{and} \bibinfo{author}{\bibfnamefont{W.~D.}
  \bibnamefont{Phillips}}, \bibinfo{journal}{Phys.\ Rev.\ A}
  \textbf{\bibinfo{volume}{67}} (\bibinfo{year}{2003}).
  
\bibitem[{\citenamefont{Rey et~al.}(2003)\citenamefont{Rey, Hu,
      Calzetta, Roura, and Clark}}]{anamaria}
  \bibinfo{author}{\bibfnamefont{A.~M.} \bibnamefont{Rey}},
  \bibinfo{author}{\bibfnamefont{B.~L.} \bibnamefont{Hu}},
  \bibinfo{author}{\bibfnamefont{E.}~\bibnamefont{Calzetta}},
  \bibinfo{author}{\bibfnamefont{A.}~\bibnamefont{Roura}},
  \bibnamefont{and}
  \bibinfo{author}{\bibfnamefont{C.}~\bibnamefont{Clark}},
  \bibinfo{journal}{Phys. \ Rev. A} \textbf{\bibinfo{volume}{69}},
  \bibinfo{number}{033610} (\bibinfo{year}{2004}).

\bibitem[{\citenamefont{Porter and
  Cvitanovi\'c}(2004{\natexlab{a}})}]{mapbecprl}
\bibinfo{author}{\bibfnamefont{M.~A.} \bibnamefont{Porter}} \bibnamefont{and}
  \bibinfo{author}{\bibfnamefont{P.}~\bibnamefont{Cvitanovi\'c}},
  \bibinfo{journal}{Phys.\ Rev.\ E} \textbf{\bibinfo{volume}{69}}
  (\bibinfo{year}{2004}{\natexlab{a}}), \bibinfo{note}{ar{X}iv:
  nlin.CD/0307032}.

\bibitem[{\citenamefont{Porter and Kevrekidis}(2004)}]{super}
\bibinfo{author}{\bibfnamefont{M.~A.} \bibnamefont{Porter}} \bibnamefont{and}
  \bibinfo{author}{\bibfnamefont{P.~G.} \bibnamefont{Kevrekidis}}
  (\bibinfo{year}{2004}), \bibinfo{note}{arXiv: nlin.PS/0406063}.

\bibitem[{\citenamefont{Bronski
  et~al.}(2001{\natexlab{a}})\citenamefont{Bronski, Carr, Deconinck, and
  Kutz}}]{bronski}
\bibinfo{author}{\bibfnamefont{J.~C.} \bibnamefont{Bronski}},
  \bibinfo{author}{\bibfnamefont{L.~D.} \bibnamefont{Carr}},
  \bibinfo{author}{\bibfnamefont{B.}~\bibnamefont{Deconinck}},
  \bibnamefont{and} \bibinfo{author}{\bibfnamefont{J.~N.} \bibnamefont{Kutz}},
  \bibinfo{journal}{Phys.\ Rev.\ Lett.} \textbf{\bibinfo{volume}{86}},
  \bibinfo{pages}{1402} (\bibinfo{year}{2001}{\natexlab{a}}).

\bibitem[{\citenamefont{Kevrekidis
  et~al.}(2003{\natexlab{b}})\citenamefont{Kevrekidis, Carretero-Gonz{\'a}lez,
  Theocharis, Frantzeskakis, and Malomed}}]{pk1}
\bibinfo{author}{\bibfnamefont{P.}~\bibnamefont{Kevrekidis}},
  \bibinfo{author}{\bibfnamefont{R.}~\bibnamefont{Carretero-Gonz{\'a}lez}},
  \bibinfo{author}{\bibfnamefont{G.}~\bibnamefont{Theocharis}},
  \bibinfo{author}{\bibfnamefont{D.}~\bibnamefont{Frantzeskakis}},
  \bibnamefont{and} \bibinfo{author}{\bibfnamefont{B.}~\bibnamefont{Malomed}},
  \bibinfo{journal}{J.\ Phys. B} \textbf{\bibinfo{volume}{36}},
  \bibinfo{pages}{3467} (\bibinfo{year}{2003}{\natexlab{b}}).

\bibitem[{\citenamefont{Kevrekidis
  et~al.}(2003{\natexlab{c}})\citenamefont{Kevrekidis, Frantzeskakis, Malomed,
  Bishop, and Kevrekidis}}]{pk2}
\bibinfo{author}{\bibfnamefont{P.}~\bibnamefont{Kevrekidis}},
  \bibinfo{author}{\bibfnamefont{D.}~\bibnamefont{Frantzeskakis}},
  \bibinfo{author}{\bibfnamefont{B.}~\bibnamefont{Malomed}},
  \bibinfo{author}{\bibfnamefont{A.}~\bibnamefont{Bishop}}, \bibnamefont{and}
  \bibinfo{author}{\bibfnamefont{I.}~\bibnamefont{Kevrekidis}},
  \bibinfo{journal}{New J. Phys.} \textbf{\bibinfo{volume}{5}},
  \bibinfo{pages}{64.1} (\bibinfo{year}{2003}{\natexlab{c}}).

\bibitem[{\citenamefont{Kevrekidis
  et~al.}(2003{\natexlab{d}})\citenamefont{Kevrekidis, Carretero-Gonz{\'a}lez,
  Theocharis, Frantzeskakis, and Malomed}}]{pk3}
\bibinfo{author}{\bibfnamefont{P.}~\bibnamefont{Kevrekidis}},
  \bibinfo{author}{\bibfnamefont{R.}~\bibnamefont{Carretero-Gonz{\'a}lez}},
  \bibinfo{author}{\bibfnamefont{G.}~\bibnamefont{Theocharis}},
  \bibinfo{author}{\bibfnamefont{D.}~\bibnamefont{Frantzeskakis}},
  \bibnamefont{and} \bibinfo{author}{\bibfnamefont{B.}~\bibnamefont{Malomed}},
  \bibinfo{journal}{Phys. Rev. A} \textbf{\bibinfo{volume}{68}},
  \bibinfo{pages}{035602} (\bibinfo{year}{2003}{\natexlab{d}}).

\bibitem[{\citenamefont{Orzel et~al.}(2001{\natexlab{b}})\citenamefont{Orzel,
  Tuchman, Fenselau, Yasuda, and Kasevich}}]{squeeze}
\bibinfo{author}{\bibfnamefont{C.}~\bibnamefont{Orzel}},
  \bibinfo{author}{\bibfnamefont{A.~K.} \bibnamefont{Tuchman}},
  \bibinfo{author}{\bibfnamefont{M.~L.} \bibnamefont{Fenselau}},
  \bibinfo{author}{\bibfnamefont{M.}~\bibnamefont{Yasuda}}, \bibnamefont{and}
  \bibinfo{author}{\bibfnamefont{M.~A.} \bibnamefont{Kasevich}},
  \bibinfo{journal}{Science} \textbf{\bibinfo{volume}{291}},
  \bibinfo{pages}{2386} (\bibinfo{year}{2001}{\natexlab{b}}).

\bibitem[{\citenamefont{Smerzi et~al.}(2002)\citenamefont{Smerzi, Trombettoni,
  Kevrekidis, and Bishop}}]{smerzi}
\bibinfo{author}{\bibfnamefont{A.}~\bibnamefont{Smerzi}},
  \bibinfo{author}{\bibfnamefont{A.}~\bibnamefont{Trombettoni}},
  \bibinfo{author}{\bibfnamefont{P.~G.} \bibnamefont{Kevrekidis}},
  \bibnamefont{and} \bibinfo{author}{\bibfnamefont{A.~R.}
  \bibnamefont{Bishop}}, \bibinfo{journal}{Phys.\ Rev.\ Lett.}
  \textbf{\bibinfo{volume}{89}}, \bibinfo{pages}{170402}
  (\bibinfo{year}{2002}).

\bibitem[{\citenamefont{Cataliotti et~al.}(2003)\citenamefont{Cataliotti,
  Fallani, Ferlaino, Fort, Maddaloni, and Inguscio}}]{cata}
\bibinfo{author}{\bibfnamefont{F.~S.} \bibnamefont{Cataliotti}},
  \bibinfo{author}{\bibfnamefont{L.}~\bibnamefont{Fallani}},
  \bibinfo{author}{\bibfnamefont{F.}~\bibnamefont{Ferlaino}},
  \bibinfo{author}{\bibfnamefont{C.}~\bibnamefont{Fort}},
  \bibinfo{author}{\bibfnamefont{P.}~\bibnamefont{Maddaloni}},
  \bibnamefont{and} \bibinfo{author}{\bibfnamefont{M.}~\bibnamefont{Inguscio}},
  \bibinfo{journal}{New J.\ Phys.} \textbf{\bibinfo{volume}{5}},
  \bibinfo{pages}{71.1} (\bibinfo{year}{2003}).

\bibitem[{\citenamefont{Porto et~al.}(2003)\citenamefont{Porto, Rolston,
  {Laburthe Tolra}, Williams, and Phillips}}]{porto}
\bibinfo{author}{\bibfnamefont{J.~V.} \bibnamefont{Porto}},
  \bibinfo{author}{\bibfnamefont{S.}~\bibnamefont{Rolston}},
  \bibinfo{author}{\bibfnamefont{B.}~\bibnamefont{{Laburthe Tolra}}},
  \bibinfo{author}{\bibfnamefont{C.~J.} \bibnamefont{Williams}},
  \bibnamefont{and} \bibinfo{author}{\bibfnamefont{W.~D.}
  \bibnamefont{Phillips}}, \bibinfo{journal}{Philosophical Transactions:
  Mathematical, Physical \& Engineering Sciences}
  \textbf{\bibinfo{volume}{361}}, \bibinfo{pages}{1417} (\bibinfo{year}{2003}).

\bibitem[{\citenamefont{Vollbrecht et~al.}(2004)\citenamefont{Vollbrecht,
  Solano, and Cirac}}]{voll}
\bibinfo{author}{\bibfnamefont{K.~G.~H.} \bibnamefont{Vollbrecht}},
  \bibinfo{author}{\bibfnamefont{E.}~\bibnamefont{Solano}}, \bibnamefont{and}
  \bibinfo{author}{\bibfnamefont{J.~L.} \bibnamefont{Cirac}}
  (\bibinfo{year}{2004}), \bibinfo{note}{arXiv: quant-ph/0405014}.

\bibitem[{\citenamefont{Porter and Cvitanovi\'c}(2004{\natexlab{b}})}]{mapbec}
\bibinfo{author}{\bibfnamefont{M.~A.} \bibnamefont{Porter}} \bibnamefont{and}
  \bibinfo{author}{\bibfnamefont{P.}~\bibnamefont{Cvitanovi\'c}},
  \bibinfo{journal}{Chaos} \textbf{\bibinfo{volume}{14}}, \bibinfo{pages}{739} 
(\bibinfo{year}{2004}{\natexlab{b}}),

\bibitem[{\citenamefont{Bronski
  et~al.}(2001{\natexlab{b}})\citenamefont{Bronski, Carr, Carretero-Gonz\'alez,
  Deconinck, Kutz, and Promislow}}]{bronskiatt}
\bibinfo{author}{\bibfnamefont{J.~C.} \bibnamefont{Bronski}},
  \bibinfo{author}{\bibfnamefont{L.~D.} \bibnamefont{Carr}},
  \bibinfo{author}{\bibfnamefont{R.}~\bibnamefont{Carretero-Gonz\'alez}},
  \bibinfo{author}{\bibfnamefont{B.}~\bibnamefont{Deconinck}},
  \bibinfo{author}{\bibfnamefont{J.~N.} \bibnamefont{Kutz}}, \bibnamefont{and}
  \bibinfo{author}{\bibfnamefont{K.}~\bibnamefont{Promislow}},
  \bibinfo{journal}{Phys.\ Rev.\ E} \textbf{\bibinfo{volume}{64}}
  (\bibinfo{year}{2001}{\natexlab{b}}).

\bibitem[{\citenamefont{Bronski
  et~al.}(2001{\natexlab{c}})\citenamefont{Bronski, Carr, Deconinck, Kutz, and
  Promislow}}]{bronskirep}
\bibinfo{author}{\bibfnamefont{J.~C.} \bibnamefont{Bronski}},
  \bibinfo{author}{\bibfnamefont{L.~D.} \bibnamefont{Carr}},
  \bibinfo{author}{\bibfnamefont{B.}~\bibnamefont{Deconinck}},
  \bibinfo{author}{\bibfnamefont{J.~N.} \bibnamefont{Kutz}}, \bibnamefont{and}
  \bibinfo{author}{\bibfnamefont{K.}~\bibnamefont{Promislow}},
  \bibinfo{journal}{Phys.\ Rev.\ E} \textbf{\bibinfo{volume}{63}}
  (\bibinfo{year}{2001}{\natexlab{c}}).

\bibitem[{\citenamefont{Alfimov
  et~al.}(2002{\natexlab{b}})\citenamefont{Alfimov, Konotop, and
  Salerno}}]{alf2}
\bibinfo{author}{\bibfnamefont{G.~L.} \bibnamefont{Alfimov}},
  \bibinfo{author}{\bibfnamefont{V.~V.} \bibnamefont{Konotop}},
  \bibnamefont{and} \bibinfo{author}{\bibfnamefont{M.}~\bibnamefont{Salerno}},
  \bibinfo{journal}{Europhys.\ Lett.} \textbf{\bibinfo{volume}{58}},
  \bibinfo{pages}{7} (\bibinfo{year}{2002}{\natexlab{b}}).

\bibitem[{\citenamefont{Louis et~al.}(2003)\citenamefont{Louis, Ostrovskaya,
  Savage, and Kivshar}}]{band}
\bibinfo{author}{\bibfnamefont{P.~J.~Y.} \bibnamefont{Louis}},
  \bibinfo{author}{\bibfnamefont{E.~A.} \bibnamefont{Ostrovskaya}},
  \bibinfo{author}{\bibfnamefont{C.~M.} \bibnamefont{Savage}},
  \bibnamefont{and} \bibinfo{author}{\bibfnamefont{Y.~S.}
  \bibnamefont{Kivshar}}, \bibinfo{journal}{Phys.\ Rev.\ A}
  \textbf{\bibinfo{volume}{67}} (\bibinfo{year}{2003}).

\bibitem[{\citenamefont{Konotop and Salerno}(2002)}]{kon}
\bibinfo{author}{\bibfnamefont{V.~V.} \bibnamefont{Konotop}} \bibnamefont{and}
  \bibinfo{author}{\bibfnamefont{M.}~\bibnamefont{Salerno}},
  \bibinfo{journal}{Phys.\ Rev.\ A} \textbf{\bibinfo{volume}{65}}
  (\bibinfo{year}{2002}).

\bibitem[{\citenamefont{Sakaguchi and Malomed}(2004)}]{sakaguchi}
\bibinfo{author}{\bibfnamefont{H.}~\bibnamefont{Sakaguchi}} \bibnamefont{and}
  \bibinfo{author}{\bibfnamefont{B.}~\bibnamefont{Malomed}},
  \bibinfo{journal}{J. Phys. B} \textbf{\bibinfo{volume}{37}},
  \bibinfo{pages}{1443} (\bibinfo{year}{2004}).

\bibitem[{\citenamefont{Abdullaev et~al.}(2001)\citenamefont{Abdullaev,
  Baizakov, Darmanyan, Konotop, and Salerno}}]{abd}
\bibinfo{author}{\bibfnamefont{F.}~\bibnamefont{Abdullaev}},
  \bibinfo{author}{\bibfnamefont{B.~B.} \bibnamefont{Baizakov}},
  \bibinfo{author}{\bibfnamefont{S.~A.} \bibnamefont{Darmanyan}},
  \bibinfo{author}{\bibfnamefont{V.~V.} \bibnamefont{Konotop}},
  \bibnamefont{and} \bibinfo{author}{\bibfnamefont{M.}~\bibnamefont{Salerno}},
  \bibinfo{journal}{Phys.\ Rev.\ A} \textbf{\bibinfo{volume}{64}}
  (\bibinfo{year}{2001}).

\bibitem[{\citenamefont{Kevrekidis and Frantzeskakis}(2004)}]{kef}
\bibinfo{author}{\bibfnamefont{P.~G.} \bibnamefont{Kevrekidis}}
  \bibnamefont{and} \bibinfo{author}{\bibfnamefont{D.~J.}
  \bibnamefont{Frantzeskakis}}, \bibinfo{journal}{Modern Phys.\ Lett.\ B}
  \textbf{\bibinfo{volume}{18}}, \bibinfo{pages}{173} (\bibinfo{year}{2004}).

\bibitem[{\citenamefont{Ashcroft and Mermin}(1976)}]{ashcroft}
\bibinfo{author}{\bibfnamefont{N.~W.} \bibnamefont{Ashcroft}} \bibnamefont{and}
  \bibinfo{author}{\bibfnamefont{N.~D.} \bibnamefont{Mermin}},
  \emph{\bibinfo{title}{Solid State Physics}}
  (\bibinfo{publisher}{Brooks/Cole}, \bibinfo{address}{Australia},
  \bibinfo{year}{1976}).

\bibitem[{\citenamefont{Trombettoni and Smerzi}(2001)}]{smer}
\bibinfo{author}{\bibfnamefont{A.}~\bibnamefont{Trombettoni}} \bibnamefont{and}
  \bibinfo{author}{\bibfnamefont{A.}~\bibnamefont{Smerzi}},
  \bibinfo{journal}{Phys.\ Rev.\ Lett.} \textbf{\bibinfo{volume}{86}},
  \bibinfo{pages}{2353} (\bibinfo{year}{2001}).

\bibitem[{\citenamefont{Strecker et~al.}(2002)\citenamefont{Strecker,
  Partridge, Truscott, and Hulet}}]{Strecker:02}
\bibinfo{author}{\bibfnamefont{K.~E.} \bibnamefont{Strecker}},
  \bibinfo{author}{\bibfnamefont{G.~B.} \bibnamefont{Partridge}},
  \bibinfo{author}{\bibfnamefont{A.~G.} \bibnamefont{Truscott}},
  \bibnamefont{and} \bibinfo{author}{\bibfnamefont{R.~G.} \bibnamefont{Hulet}},
  \bibinfo{journal}{Nature} \textbf{\bibinfo{volume}{417}},
  \bibinfo{pages}{150} (\bibinfo{year}{2002}).

\bibitem[{\citenamefont{Khawaja et~al.}(2002)\citenamefont{Khawaja, Stoof,
  Hulet, Strecker, and Partridge}}]{Khawaja:02}
\bibinfo{author}{\bibfnamefont{U.~A.} \bibnamefont{Khawaja}},
  \bibinfo{author}{\bibfnamefont{H.~T.~C.} \bibnamefont{Stoof}},
  \bibinfo{author}{\bibfnamefont{R.~G.} \bibnamefont{Hulet}},
  \bibinfo{author}{\bibfnamefont{K.~E.} \bibnamefont{Strecker}},
  \bibnamefont{and} \bibinfo{author}{\bibfnamefont{G.~B.}
  \bibnamefont{Partridge}}, \bibinfo{journal}{Phys.\ Rev.\ Lett.}
  \textbf{\bibinfo{volume}{89}}, \bibinfo{pages}{200404}
  (\bibinfo{year}{2002}).

\bibitem[{\citenamefont{Strecker et~al.}(2003)\citenamefont{Strecker,
  Partridge, Truscott, and Hulet}}]{Strecker:03}
\bibinfo{author}{\bibfnamefont{K.~E.} \bibnamefont{Strecker}},
  \bibinfo{author}{\bibfnamefont{G.~B.} \bibnamefont{Partridge}},
  \bibinfo{author}{\bibfnamefont{A.~G.} \bibnamefont{Truscott}},
  \bibnamefont{and} \bibinfo{author}{\bibfnamefont{R.~G.} \bibnamefont{Hulet}},
  \bibinfo{journal}{New J.\ Phys.} \textbf{\bibinfo{volume}{5}},
  \bibinfo{pages}{73.1} (\bibinfo{year}{2003}).

\bibitem[{\citenamefont{Khaykovich et~al.}(2002)\citenamefont{Khaykovich,
  Schreck, Ferrari, Bourdel, Cubizolles, Carr, Castin, and Salomon}}]{Carr:02}
\bibinfo{author}{\bibfnamefont{L.}~\bibnamefont{Khaykovich}},
  \bibinfo{author}{\bibfnamefont{F.}~\bibnamefont{Schreck}},
  \bibinfo{author}{\bibfnamefont{G.}~\bibnamefont{Ferrari}},
  \bibinfo{author}{\bibfnamefont{T.}~\bibnamefont{Bourdel}},
  \bibinfo{author}{\bibfnamefont{J.}~\bibnamefont{Cubizolles}},
  \bibinfo{author}{\bibfnamefont{L.~D.} \bibnamefont{Carr}},
  \bibinfo{author}{\bibfnamefont{Y.}~\bibnamefont{Castin}}, \bibnamefont{and}
  \bibinfo{author}{\bibfnamefont{C.}~\bibnamefont{Salomon}},
  \bibinfo{journal}{Science} \textbf{\bibinfo{volume}{296}},
  \bibinfo{pages}{1290} (\bibinfo{year}{2002}).

\bibitem[{\citenamefont{Toda}(1989)}]{Toda:book}
\bibinfo{author}{\bibfnamefont{M.}~\bibnamefont{Toda}},
  \emph{\bibinfo{title}{Theory of Nonlinear Lattices.}}
  (\bibinfo{publisher}{Springer-Verlag}, \bibinfo{year}{1989}),
  \bibinfo{edition}{2nd} ed.

\bibitem[{\citenamefont{Carretero-Gonz\'{a}lez and
  Promislow}(2002)}]{promislow}
\bibinfo{author}{\bibfnamefont{R.}~\bibnamefont{Carretero-Gonz\'{a}lez}}
  \bibnamefont{and}
  \bibinfo{author}{\bibfnamefont{K.}~\bibnamefont{Promislow}},
  \bibinfo{journal}{Phys.\ Rev.\ A} \textbf{\bibinfo{volume}{66}}
  (\bibinfo{year}{2002}).

\bibitem[{\citenamefont{Carretero-Gonz\'{a}lez and Promislow}(2004)}]{ric2}
\bibinfo{author}{\bibfnamefont{R.}~\bibnamefont{Carretero-Gonz\'{a}lez}}
  \bibnamefont{and} \bibinfo{author}{\bibfnamefont{K.}~\bibnamefont{Promislow}}
  (\bibinfo{year}{2004}), \bibinfo{note}{in preparation}.

\bibitem[{\citenamefont{Malomed}(2002)}]{Progress}
\bibinfo{author}{\bibfnamefont{B.}~\bibnamefont{Malomed}},
  \bibinfo{journal}{Progress in Optics} \textbf{\bibinfo{volume}{43}},
  \bibinfo{pages}{69} (\bibinfo{year}{2002}).

\bibitem[{\citenamefont{Karpman and
  Solov'ev}(1981{\natexlab{a}})}]{Karpman:81a}
\bibinfo{author}{\bibfnamefont{V.~I.} \bibnamefont{Karpman}} \bibnamefont{and}
  \bibinfo{author}{\bibfnamefont{V.~V.} \bibnamefont{Solov'ev}},
  \bibinfo{journal}{Physica D} \textbf{\bibinfo{volume}{3}},
  \bibinfo{pages}{142} (\bibinfo{year}{1981}{\natexlab{a}}).

\bibitem[{\citenamefont{Karpman and
  Solov'ev}(1981{\natexlab{b}})}]{Karpman:81b}
\bibinfo{author}{\bibfnamefont{V.~I.} \bibnamefont{Karpman}} \bibnamefont{and}
  \bibinfo{author}{\bibfnamefont{V.~V.} \bibnamefont{Solov'ev}},
  \bibinfo{journal}{Physica D} \textbf{\bibinfo{volume}{3}},
  \bibinfo{pages}{487} (\bibinfo{year}{1981}{\natexlab{b}}).

\bibitem[{\citenamefont{Gerdjikov et~al.}(1997)\citenamefont{Gerdjikov, Uzunov,
  Estaviev, and Diankov}}]{Gerdjikov:97}
\bibinfo{author}{\bibfnamefont{V.~S.} \bibnamefont{Gerdjikov}},
  \bibinfo{author}{\bibfnamefont{I.~M.} \bibnamefont{Uzunov}},
  \bibinfo{author}{\bibfnamefont{E.~G.} \bibnamefont{Estaviev}},
  \bibnamefont{and} \bibinfo{author}{\bibfnamefont{G.~L.}
  \bibnamefont{Diankov}}, \bibinfo{journal}{Phys.\ Rev.\ E}
  \textbf{\bibinfo{volume}{55}}, \bibinfo{pages}{6039} (\bibinfo{year}{1997}).

\bibitem[{\citenamefont{Arnold}(1998)}]{Arnold:98}
\bibinfo{author}{\bibfnamefont{J.~M.} \bibnamefont{Arnold}},
  \bibinfo{journal}{J.\ Opt.\ Soc.\ Am.\ A} \textbf{\bibinfo{volume}{15}},
  \bibinfo{pages}{1450} (\bibinfo{year}{1998}).

\bibitem[{\citenamefont{Arnold}(1999)}]{Arnold:99}
\bibinfo{author}{\bibfnamefont{J.~M.} \bibnamefont{Arnold}},
  \bibinfo{journal}{Phys.\ Rev.\ E} \textbf{\bibinfo{volume}{60}},
  \bibinfo{pages}{979} (\bibinfo{year}{1999}).

\bibitem[{\citenamefont{Kevrekidis et~al.}(2001)\citenamefont{Kevrekidis,
  Rasmussen, and Bishop}}]{ijmpbrev}
\bibinfo{author}{\bibfnamefont{P.}~\bibnamefont{Kevrekidis}},
  \bibinfo{author}{\bibfnamefont{K.}~\bibnamefont{Rasmussen}},
  \bibnamefont{and} \bibinfo{author}{\bibfnamefont{A.}~\bibnamefont{Bishop}},
  \bibinfo{journal}{Int. J. Mod. Phys. B} \textbf{\bibinfo{volume}{15}},
  \bibinfo{pages}{2833} (\bibinfo{year}{2001}).

\bibitem[{\citenamefont{Malomed and Kevrekidis}(2001)}]{borya1}
\bibinfo{author}{\bibfnamefont{B.}~\bibnamefont{Malomed}} \bibnamefont{and}
  \bibinfo{author}{\bibfnamefont{P.}~\bibnamefont{Kevrekidis}},
  \bibinfo{journal}{Phys.\ Rev.\ E} \textbf{\bibinfo{volume}{64}},
  \bibinfo{pages}{026601} (\bibinfo{year}{2001}).

\bibitem[{\citenamefont{Kevrekidis et~al.}(2004)\citenamefont{Kevrekidis,
  Malomed, Frantzeskakis, and Carretero-Gonz{\'a}lez}}]{borya2}
\bibinfo{author}{\bibfnamefont{P.}~\bibnamefont{Kevrekidis}},
  \bibinfo{author}{\bibfnamefont{B.}~\bibnamefont{Malomed}},
  \bibinfo{author}{\bibfnamefont{D.}~\bibnamefont{Frantzeskakis}},
  \bibnamefont{and}
  \bibinfo{author}{\bibfnamefont{R.}~\bibnamefont{Carretero-Gonz{\'a}lez}},
  \bibinfo{journal}{Phys.\ Rev.\ Lett.} \textbf{\bibinfo{volume}{93}},
  \bibinfo{pages}{080403} (\bibinfo{year}{2004}).

\end{thebibliography}

\end{document}